\documentclass[prd,twocolumn,preprintnumbers,
amsmath,amssymb,aps,nofootinbib,superscriptaddress]{revtex4-1}
\usepackage[pdftex]{graphicx}
\usepackage{dcolumn}
\usepackage{bm}
\usepackage{cases}
\usepackage{natbib}
\usepackage[pdftex]{hyperref}
\usepackage{color}
\definecolor{phthaloblue}{rgb}{0.0, 0.06, 0.54}
\definecolor{bluscuro}{rgb}{0.15, 0.2, .85}
\definecolor{rossos}{cmyk}{0,1,1,0.55}
\definecolor{bluchiaro}{cmyk}{1,.3,0.,0.1}
\hypersetup{
    colorlinks=true,
    linkcolor=bluscuro,
    citecolor=phthaloblue,
    filecolor=phthaloblue,
    urlcolor=phthaloblue,
    }
\makeatletter \def\@eqnnum{{\normalsize \normalcolor (\theequation)}}  
\makeatother %
\newcommand{\para}[1]{\par\vspace{2mm}\noindent\textbf{\emph{{#1}}.---}}

\begin{document}
\title{Lorentzian path integral for quantum tunneling and 
WKB approximation \\for wave-function}
\author{Hiroki Matsui}
\email{hiroki.matsui@yukawa.kyoto-u.ac.jp}
\affiliation{Center for Gravitational Physics, Yukawa Institute for Theoretical Physics, \\ Kyoto University, 606-8502, Kyoto, Japan}

\begin{abstract}
Recently, the Lorentzian path integral formulation using the Picard-Lefschetz theory has attracted much attention in quantum cosmology. 
In this paper, we analyze the tunneling 
amplitude in quantum mechanics by using the Lorentzian 
Picard-Lefschetz formulation
and compare it with the WKB analysis of the conventional 
Schr\"{o}dinger equation. We show that the Picard-Lefschetz Lorentzian formulation 
is consistent with the WKB approximation for wave-function and the Euclidean path integral formulation utilizing the solutions of the 
Euclidean constraint equation.
We also consider some problems of this Lorentzian Picard-Lefschetz formulation and
discuss a simpler semiclassical approximation of the Lorentzian path integral
without integrating the lapse function.
\end{abstract}

\maketitle
\flushbottom
\allowdisplaybreaks[1]

%%%%%%%%%%%%%%%%%%%%%%%%%%%%%%%%%%%%
\section{Intorduction}
%%%%%%%%%%%%%%%%%%%%%%%%%%%%%%%%%%%%
Wave function distinguishes quantum and classical picture of physical systems
and can be exactly calculated by solving the Schr\"{o}dinger equation
for quantum mechanics (QM).
In particular, quantum tunneling is one of the 
most important consequences of the wave function.
The wave function inside the potential barrier seeps out the barrier even if the kinetic energy is lower than the potential energy. As a result, the non-zero probability occurs outside the potential in the quantum system even if it is bound by the potential barrier in the classical system.
Hence, quantum tunneling is one of the most important phenomena 
to describe the quantum nature of the system.

The Feynman's path integral~\cite{Feynman:1948ur} 
is a standard formulation for QM and quantum field theory (QFT) 
which is equivalent with Schr\"{o}dinger equation
and defines quantum transition amplitude which is given by the integral over all paths 
weighted by the factor $e^{iS[x]/\hbar}$.
In the path integral formulation, the transition amplitude from an initial state $x(t_i)=x_i$ to a final state $x(t_f)=x_f$ is written by the functional integral,
\begin{align} \label{eqn:path-integral}
K(x_{f};x_{i})&=\langle x_f,t_f \mid x_i,t_i \rangle \notag \\ &= 
\int_{x(t_{i})=x_{i}}^{x(t_{f})=x_{f}}  \mathcal{D}x(t) \, \exp\left(\frac{iS[x]}{\hbar}\right),
\end{align}
where we consider a unit mass particle whose 
the action $S[x]$ is written by
\begin{equation} \label{eqn:action1}
S[x]=\int \mathrm{d}t \left( \frac{1}{2}\dot{x}^2 - V(x) \right) \,.
\end{equation}

In a semi-classical regime, the associated path integral can be 
given by the saddle point approximation which is dominated by the path $\delta S[x]/\delta x \approx 0$.
In particular, to describe the quantum tunneling in Feynman's formulation 
the Euclidean path integral method~\cite{Coleman:1977py} is used. 
Performing the Wick rotation $\tau = i t$ to Euclidean time,
the dominant field configuration is given by solutions of the
Euclidean equations of motion imposed by boundary conditions.
The instanton constructed by the half-bounce Euclidean 
solutions going from $x_i$ to $x_f$~\cite{Polyakov:1976fu} describes 
the tunneling event across a degenerate potential and the splitting energy.
On the other hand, the bounce constructed by the bouncing 
 Euclidean solutions with $x_i=x_f$ gives the vacuum decay ratio 
 from the local-minimum false vacuum~\cite{Coleman:1977py}.
 The Euclidean instanton method is useful tool for QFT 
\cite{Belavin:1975fg,Polyakov:1976fu,Coleman:1977py,Callan:1977pt} and even 
the gravity~\cite{Coleman:1980aw}.

However, the method is conceptually less straightforward. When a particle tunnels through a potential barrier, we can accurately calculate the tunneling probability by using the Euclidean action $S_E[x]$ with the instanton solution, 
and in fact, it works well. 
However, when the particle is moving outside the potential, 
one needs to use the Lorentzian or real-time path integral so that the instanton 
formulation lacks unity and it is unclear why this method works.
In this perspective, the quantum tunneling of the Lorentzian or real-time
path integrals has been studied in recent years~\cite{Levkov:2004ij,
Bender:2008fr,Bender:2009zza,Bender:2009jg,Bender:2010nu,
Dumlu:2011cc,Turok:2013dfa,Tanizaki:2014xba,Cherman:2014sba,
Behtash:2015zha,Behtash:2015loa,
Ilderton:2015qda,Bramberger:2016yog,Ai:2019fri}, where complex instanton solutions are considered. 
It has been argued in Ref.~\cite{Cherman:2014sba,Ai:2019fri} that the 
Euclidean-time instanton solutions for a rotated time $t=\tau e^{-i\alpha}$ close to 
Lorentzian-time describes something like a real-time description of 
quantum tunneling.
Besides the extensions of the instanton method, a new tunneling
approach has been proposed by Ref.~\cite{Feldbrugge:2017kzv}
for quantum cosmology where the
Lorentzian path integral includes a lapse integral 
and the saddle-point integration is performed by the Picard-Lefschetz theory
(we call this method Lorentzian Picard-Lefschetz formulation). 
Feldbrugge $\textrm{et al.}$~\cite{Feldbrugge:2017kzv} showed that 
the Lorentzian path integral reduces to 
the Vilenkin's tunneling wave function~\cite{Vilenkin:1984wp}
by perfuming the integral over a contour.
On other hand, Diaz Dorronsoro $\textrm{et al.}$
reconsidered the Lorentzian path integral 
by integrating the lapse gauge over a different
contour~\cite{DiazDorronsoro:2017hti}
and show that the Lorentzian path integral reduces to be 
the Hartle-Hawking's no boundary wave function~\cite{Hartle:1983ai}.
Both the tunneling or no-boundary wave functions can be derived as the
Wentzel-Kramers-Brillouin (WKB) solutions of the Wheeler-DeWitt equation 
in mini-superspace model~\cite{Vilenkin:1987kf,Vilenkin:1994rn,Vilenkin:1998dn}.
However, it is not fully understood why different wave functions 
can be obtained by different contours of the lapse integration in the Lorentzian path integral
of quantum gravity (QG) 
and why the gravitational amplitude using the method of steepest descents  
or saddle-point corresponds to the WKB solution 
of the Wheeler-DeWitt equation~\cite{Halliwell:1988ik,Halliwell:1989vu,Halliwell:1990tu,
Brown:1990iv,Vilenkin:2018dch,deAlwis:2018sec}
%%%%%%%%%%%%%%%%%%%%%%%%%%%%%%%
\footnote{
The perturbation issues for the tunneling or 
no-boundary wave functions in the Lorentzian path integral
have been discussed in Refs~\cite{Feldbrugge:2017fcc,Feldbrugge:2017mbc, Feldbrugge:2018gin,DiazDorronsoro:2018wro, Halliwell:2018ejl,Janssen:2019sex,deAlwis:2018sec,Vilenkin:2018dch,Vilenkin:2018oja,Bojowald:2018gdt, DiTucci:2018fdg, DiTucci:2019dji, DiTucci:2019bui}.}.
%%%%%%%%%%%%%%%%%%%%%%%%%%%%%%%

In this paper, we apply this Lorentzian path integral method 
to the tunneling amplitude of QM 
to discuss these conundrums without 
the complications associated with QG.
We will reconfirm that the conjecture of the tunneling or 
no-boundary wave functions based on the path integral of QG
holds for QM as well, and show that the path integral~\eqref{QM-amplitude} 
under the Lorentzian Picard-Lefschetz method
corresponds to the WKB wave function 
of the Schr\"{o}dinger equation.
We will provide some examples 
in Section~\ref{sec:Lorentzian-path-integral}
and Section~\ref{sec:Euclidean-path-integral}.
Furthermore, we will discuss and confirm 
the relations between the Lorentzian Picard-Lefschetz formulation,
and the WKB approximation for wave-function and the standard instanton method based on the 
Euclidean path integral. We also consider some problems of this Lorentzian Picard-Lefschetz  formulation and
discuss a simpler semiclassical approximation of the Lorentzian path integral.

The present paper is organized as follows. In
Section~\ref{sec:Lorentzian-path-integral} we
introduce the Lorentzian path integral with the Picard-Lefschetz theory
and apply this Lorentzian Picard-Lefschetz formulation to the 	
linear, harmonic oscillator, and inverted harmonic oscillator models.
In Section~\ref{sec:Euclidean-path-integral} we review the Euclidean path integral, 
instanton and WKB approximation and consider their relations to the 
Lorentzian Picard-Lefschetz formulation.
In Section~\ref{sec:Lorentzian-gravity}
we demonstrate that 
the tunneling and no-boundary wave functions derived by 
the Lorentzian Picard-Lefschetz Formulation corresponds to the WKB solution 
of the Wheeler-DeWitt equation.
In Section~\ref{sec:Lorentzian-instanton} we discuss a simpler semiclassical approximation 
method of the Lorentzian path integral without involving the lapse integral. 
Finally, in
Section~\ref{sec:Conclusion} we conclude our work.

%%%%%%%%%%%%%%%%%%%%%%%%%%%%%%%%%%%%
%%%%%%%%%%%%%%%%%%%%%%%%%%%%%%%%%%%%
\section{Lorentzian path integral with Picard-Lefschetz theory }
\label{sec:Lorentzian-path-integral} 
%%%%%%%%%%%%%%%%%%%%%%%%%%%%%%%%%%%%
%%%%%%%%%%%%%%%%%%%%%%%%%%%%%%%%%%%%

\begin{figure}[t] 
	\centering
	\includegraphics[width=0.23\textwidth]{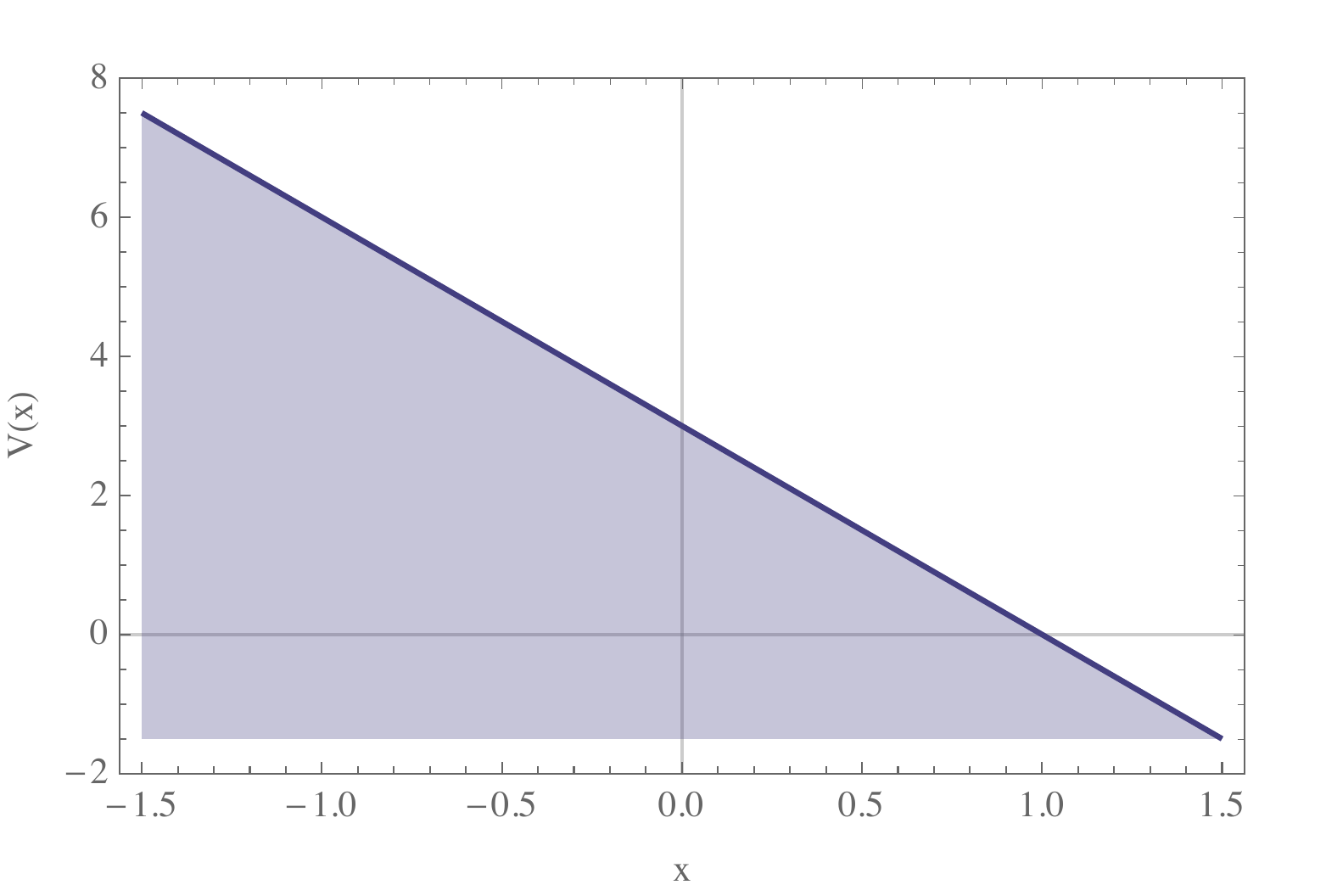}
	\includegraphics[width=0.23\textwidth]{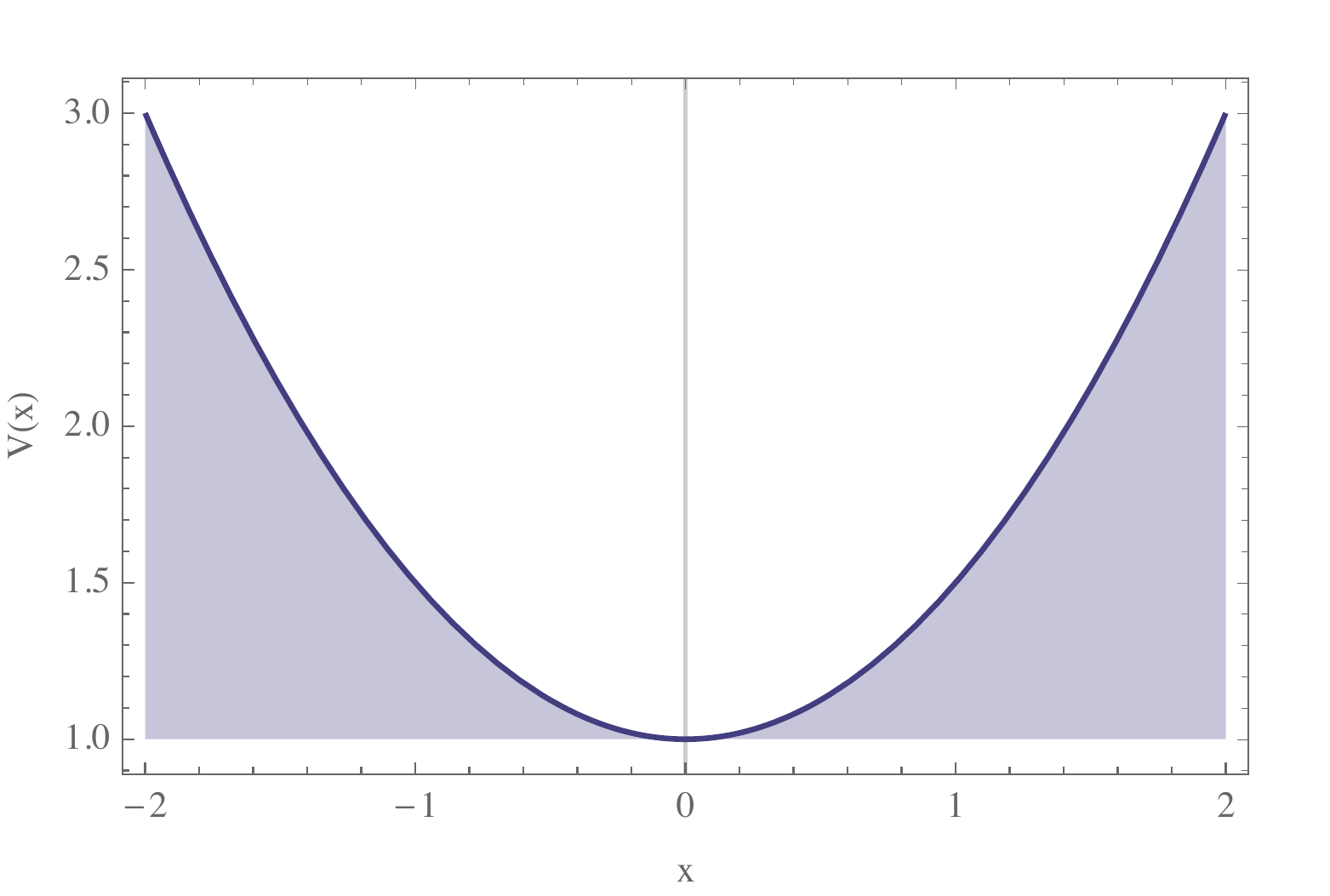}
	\includegraphics[width=0.23\textwidth]{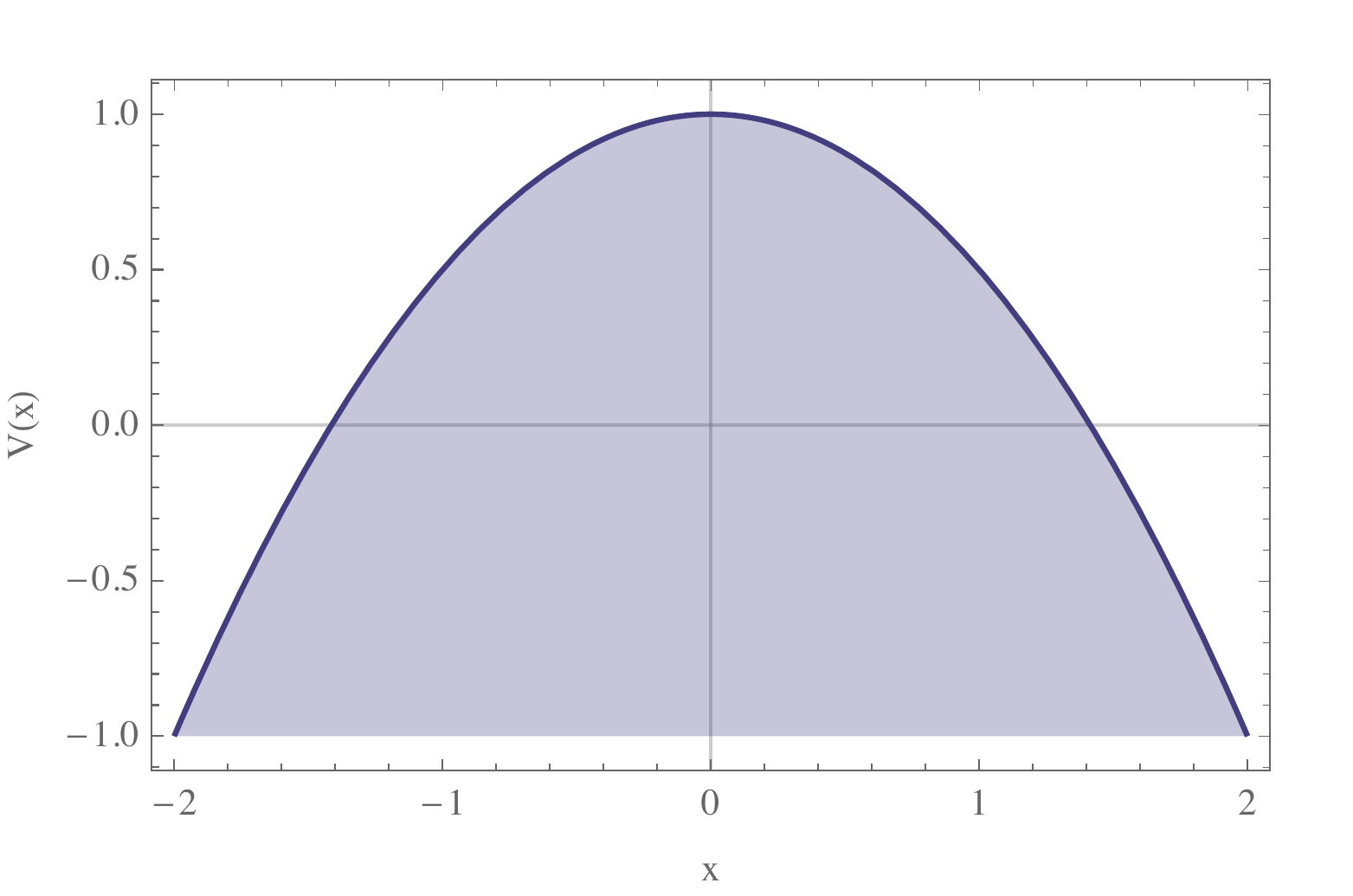}
	\includegraphics[width=0.23\textwidth]{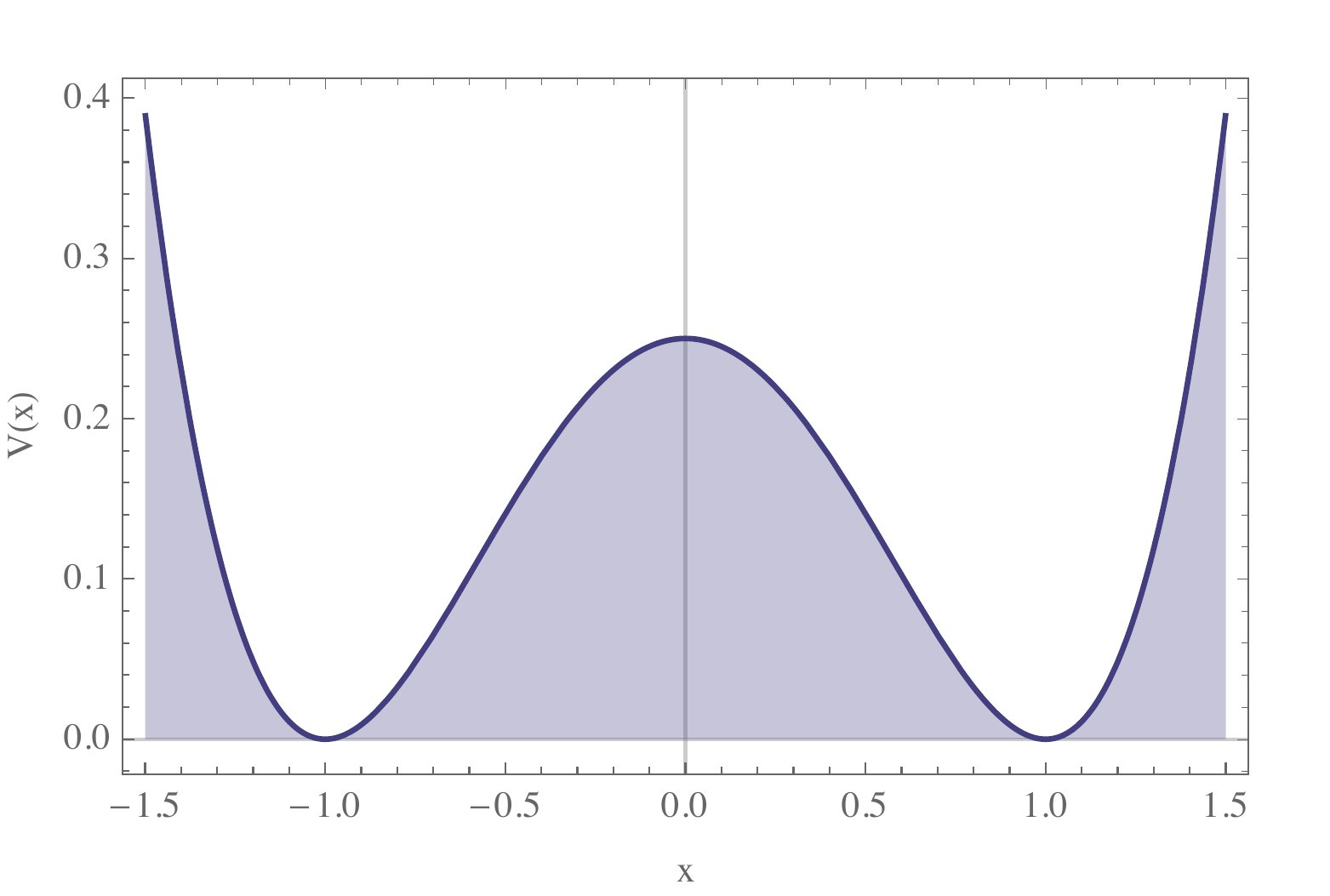}
		\caption{These figures show the potential $V(x)$ for the 	
	linear, harmonic oscillator, inverted harmonic oscillator, and double well models.
	In this paper, we consider the Lorentzian path integral for these potentials.}
	\label{fig:potential}
\end{figure}

We introduce the 
Lorentzian path integral for QM and apply
the steepest descents or saddle-point method utilizing 
the Picard-Lefschetz theory to the 
Lorentzian path integral.
The Lorentzian
path integral for QM is given by,
%%%%%%%%%%%%%%%%%%%%%%%%%%%%%%%%%%%%
\footnote{
After revising this paper, we noticed that the content of this paper 
was very similar to that of Ref~\cite{Carlitz:1984ab}. 
Based on the discussion in~\cite{Carlitz:1984ab}, 
it may be reasonable to consider the Lorentzian path integral simply as 
a complex integral representation of Green's function rather than
the extension of the Euclidean path integral. }
%%%%%%%%%%%%%%%%%%%%%%%%%%%%%%%%%%%%
\begin{align}\label{QM-amplitude}
K(x_{f};x_{i})=\int\mathcal{D} N(t) \int_{x(t_{0})=x_{0}}^{x(t_{1})=x_{1}} 
\mathcal{D}x(t) ~ \exp \left(\frac{i S[N,x]}{\hbar}\right) ~,
\end{align}
where $N(t)$ is the lapse function.
From here we fix the gauge: $N(t)=N=\textrm{const.}$.
Extending the lapse function $N$ from real $\mathcal{R}$ to complex $\mathcal{C}$
enable to consider classically prohibited evolution of the particles
where $N=1$ corresponds to moving along the real-time whereas 
$N=-i$ corresponds to the Euclidean time. 
The action $S[N,x]$ is written as 
\begin{equation}\label{QM-action}
S[N,x]=\int_{t_i}^{t_f}dt N(t)
\left( \frac{\dot{x}^{2}}{2N(t)^2}-V(x)+E\right),
\end{equation}
where $V(x)$ is the potential and $E$ is the energy of the system.
We will discuss the linear, harmonic oscillator, 
inverted harmonic oscillator and double well models for QM. 
Fig.~\ref{fig:potential} shows the potential $V(x)$ for these models.
From \eqref{QM-action} we derive the following constraint equation and 
equations of motion~\cite{deAlwis:2018sec},
\begin{align}
\delta S[x,N]/\delta{N}& =  0\ \Longrightarrow \ 
\frac{\dot{x}^{2}}{2}+N^{2}V(x)=N^{2}E\label{eq:Neqn},\\
\delta S[x,N]/\delta{x} & =  0 \ \Longrightarrow \ \ddot{x}=-N^{2}V'(x)\label{eq:xeqn},
\end{align}

%%%%%%%%%%%%%%%%%%%%%%%%%%%%%%%%%%%%%%%%%%%
%%%%%%%%%%%%%%%%%%%%%%%%%%%%%%%%%%%%%%%%%%%
%%%%%%%%%%%%%%%%%%%%%%%%%%%%%%%%%%%%%%%%%%%
%%%%%%%%%%%%%%%%%%%%%%%%%%%%%%%%%%%%%%%%%%%

It should be emphasized that 
the definition of the Lorentzian path integral~\eqref{QM-amplitude}
is not necessarily the same as the original path integral~\eqref{eqn:path-integral}.
But, there is a correspondence between these formulations 
from the definition of path integral,
\begin{align}
\langle x_f \mid 
e^{-i H (t_f-t_i)} \mid x_i \rangle = 
\int \mathcal{D}x(t) \, e^{{iS[x]}/{\hbar}},
\end{align}
as the Euclidean path integral formulation which 
is given by the standard Wick rotation $t\rightarrow -i\tau$.
The Lorentzian path integral~\eqref{QM-amplitude} is given by 
the transformation of $t\rightarrow Nt$ and can be regarded as 
a complex-time formulation of the path integral by
assuming $N$ to be complex.

Although there have been several works on such complex path integral methods~\cite{Mclaughlin:1972ws,Aoyama:1994zw},
there is no obvious choice of integration contours in the complex-time path integral.
But, as will be shown later,
we solve this problem by using the
method of steepest descents  
or saddle-point method~\cite{Halliwell:1988ik,Halliwell:1989vu,Halliwell:1990tu,
Brown:1990iv}. Moreover, the Picard-Lefschetz theory allows us to develop 
this argument more mathematically and rigorously. 
This theory provides a unique way to find a complex integration contour
based on the steepest descent path (Lefschetz thimbles $\cal J_\sigma$) 
and proceed with such oscillatory integral as~\cite{Witten:2010cx}, 
\begin{equation}
\int_\mathcal{R} \mathrm{d}x \, \exp\left(\frac{i S[x]}{\hbar}\right)
=\sum_\sigma n_\sigma \int_{\cal J_\sigma} \mathrm{d}x
\exp\left(\frac{i S[x_\sigma]}{\hbar}\right).
\end{equation} 
where $n_\sigma$ is the intersection number of the Lefschetz thimbles $\cal J_\sigma$ 
and steepest ascent path $\cal K_\sigma$.
In this section, we apply this Lorentzian 
Picard-Lefschetz method to the tunneling transition of QM.
%%%%%%%%%%%%%%%%%%%%%%%%%
%%%%%%%%%%%%%%%%%%%%%%%%%
\footnote{
This method assumes the semi-classical approximation, 
and other methods utilizing the Picard-Lefschetz theory might 
provide a complete analysis of the path integral for QM/QFT
(see e.g.,~\cite{Mou:2019tck,Mou:2019gyl,Mou:2020uuk}).}
%%%%%%%%%%%%%%%%%%%%%%%%%
%%%%%%%%%%%%%%%%%%%%%%%%%

%%%%%%%%%%%%%%%%%%%%%%%%%%%%%%%%%%%%
\subsection{Quantum tunneling with Lorentzian path integral}
%%%%%%%%%%%%%%%%%%%%%%%%%%%%%%%%%%%%

Now, we will consider the linear potential $V=V_{0}-\Lambda x$ with 
$\Lambda>0$ which corresponds to the no-boundary proposal 
of the Lorentzian path integral for QG~\cite{Feldbrugge:2017kzv}.
Thus, we have the following action,
\begin{equation}\label{action1}
S[x,N]= \int_{0}^{1}dt N
\left( \frac{\dot{x}^{2}}{2N^2}-V_{0}+\Lambda x+E\right),
\end{equation}
whose classical solution is given as
\begin{equation}\label{x-semisol}
x_{s}=\frac{\Lambda}{2}N^{2} t^{2}+\left(-\frac{1}{2}N^{2}\Lambda+x_{1}-x_{0}
\right)t+x_{0}.
\end{equation}

Following~\cite{Feldbrugge:2017kzv,Halliwell:1988ik},
we can evaluate the Lorentzian path integral~\eqref{QM-amplitude}
under the semi-classical approximation.
We assume the full solution $x(t) = x_s(t) + Q(t)$ where
$Q(t)$ is the Gaussian fluctuation around the semi-classical solution~\eqref{x-semisol}.
By substituting it for the action~\eqref{action1} and 
integrating the path integral over $Q(t)$, 
%%%%%%%%%%%%%%%%%%%%%%%%%%%%%%%%%
\footnote{
We used the following path integral formulation,
\begin{equation*} 
\int_{X[0]=0}^{X[1]=0} \mathcal{D}X(t) \exp \left(\frac{i}{\hbar}
 \int_0^1 \mathrm{d}t\, \frac{1}{2}m \dot{x}^2\right)=
\sqrt{\frac{m}{2\pi i\hbar}} \,.
\end{equation*}}
%%%%%%%%%%%%%%%%%%%%%%%%%%%%%%%%%
we have 
the following oscillatory integral,
\begin{equation}\label{o-integral}
K(x_{1};x_{0}) = \sqrt{\frac{1}{2\pi i\hbar }} \int_\mathcal{C}
\frac{\mathrm{d} N}{N^{1/2}} \exp\left(\frac{i S_0[N]}{\hbar}\right),
\end{equation}
where 
\begin{align}\label{semi-action}
&S_0[N]  =\int_{0}^{1}dt N
\left( \frac{\dot{x}_s^{2}}{2N^2}-V_{0}+\Lambda x_s+E\right)\nonumber \\
&= -\frac{\Lambda^{2}N^{3}}{24}-N\left(-\frac{\Lambda}{2}(x_{1}+x_{0})
-E+V_{0}\right)+\frac{(x_{1}-x_{0})^{2}}{2N}\,.
\end{align}
Thus, we can calculate the transition amplitude by only
performing the integration of the lapse function.
Although it is generally difficult to handle such oscillatory integrals,
the Picard-Lefschetz theory deals with such integrals.
The Picard-Lefschetz theory complexifies the variables and 
selects a complex path such that the original integral 
does not change formally via an extension of 
Cauchy's integral theorem, and especially pass the saddle points 
known as the Lefschetz thimbles $\cal J_\sigma$.

Now let us integrate the lapse $N$ integral 
along the Lefschetz thimbles ${\cal J}_\sigma$, and we obtain 
\begin{align}\label{PL-integral}
K(x_{1};x_{0}) & =  \sum_\sigma n_\sigma \sqrt{\frac{1}{2\pi i\hbar }}
\int_{{\cal J}_\sigma} \frac{\mathrm{d}N}{N^{1/2}} \exp\left(\frac{i S_0[N]}{\hbar}\right).
\end{align}
Since the lapse integral~\eqref{PL-integral} can be approximately estimated based on 
the saddle points $N_s$ and solving ${\partial S_0[N]}/{\partial N}=0$,
the saddle-points of the action $S_0[N]$ 
are given by,
\begin{equation}\label{saddle-lapse-L}
N_s=a_1\frac{\sqrt{2}}{\Lambda}
\left[(\Lambda x_{0}+E-V_{0})^{1/2}
+a_2(\Lambda x_{1}+E-V_{0})^{1/2}\right],
\end{equation}
where $a_1,a_2\in\{-1,1\}$.
The four saddle points~\eqref{saddle-lapse-L} correspond to the intersection
of the steepest descent path $\cal J_\sigma$ (Lefschetz thimbles) and steepest ascent 
path $\cal K_\sigma$ 
where $\textrm{Re}\left[iS_0\left( N \right)\right]$
decreases and increases monotonically on $\cal J_\sigma$ and $\cal K_\sigma$.
The saddle-point action $S_0[N_s]$ evaluated at $N_s$ is given by
\begin{align}\label{saddle-action-L}
S_0[N_s] = a_1 \frac{2\sqrt{2}}{3\Lambda}  \left[ \left(\Lambda x_{0}+E-V_{0}\right)^{3/2} +a_2 \left(\Lambda x_{1}+E-V_{0}\right)^{3/2} \right]\,.
\end{align}
Thus, using the saddle-point approximation we can get the following result,
\begin{align}\label{saddle}
K(x_{1};x_{0}) 
& \approx \sum_\sigma n_\sigma e^{i\theta_\sigma} 
\sqrt{\frac{1}{2\pi i\hbar }} \frac{\exp\left({i S_0[N_s]/\hbar}\right)}{N_s^{1/2}}
\nonumber\\ 
&\times \int_{{\cal J}_\sigma} \mathrm{d}R \exp \left({-\frac{1}{2\hbar}\left|\frac{\partial^2S_0[N_s]}{\partial N^2}\right|R^2} \right)\\ 
& \approx \sum_\sigma n_\sigma e^{i\theta_\sigma} \sqrt{\frac{1}{ iN_s \left|\frac{\partial^2S_0[N_s]}{\partial N^2}\right|}}  \exp\left(\frac{i S_0[N_s]}{\hbar}\right)\,,\nonumber
\end{align}
where we expand $S_0[N]$ around a saddle point $N_s$ as follows,
%%%%%%%%%%%%%%%%%%%%%%%%%
\footnote{~We introduced
$N-N_s \equiv R e^{i\theta_\sigma}$ and 
$\textrm{Arg}\left( \frac{\partial^2S_0[N]}{\partial N^2}\Bigr|_{N=N_s}\right)=\alpha$.
Thus, we get $e^{i(2\theta_\sigma+\alpha)}=i$ and $\theta_\sigma=\pi/4-\alpha/2$.
}
%%%%%%%%%%%%%%%%%%%%%%%%%
\begin{align}
\frac{iS_0[N]}{\hbar}&=\frac{iS_0[N]}{\hbar}\Bigr|_{N=N_s} 
-\frac{1}{2\hbar}\left|\frac{\partial^2S_0[N_s]}{\partial N^2}\right|R^2\nonumber\\ 
&+ \frac{i}{6\hbar}\frac{\partial^3S_0[N]}
{\partial N^3}\Bigr|_{N=N_s}(N-N_s)^3+ \dots 
\end{align}

%%%%%%%%%%%%%%%%%%%%%%%%%%%%%%%%%%%%%%%%%%%%%%%%
%%%%%%%%%%%%%%%%%%%%%%%%%%%%%%%%%%%%%%%%%%%%%%%%
\begin{figure*}[t]
	\centering
	\includegraphics[scale=0.6]{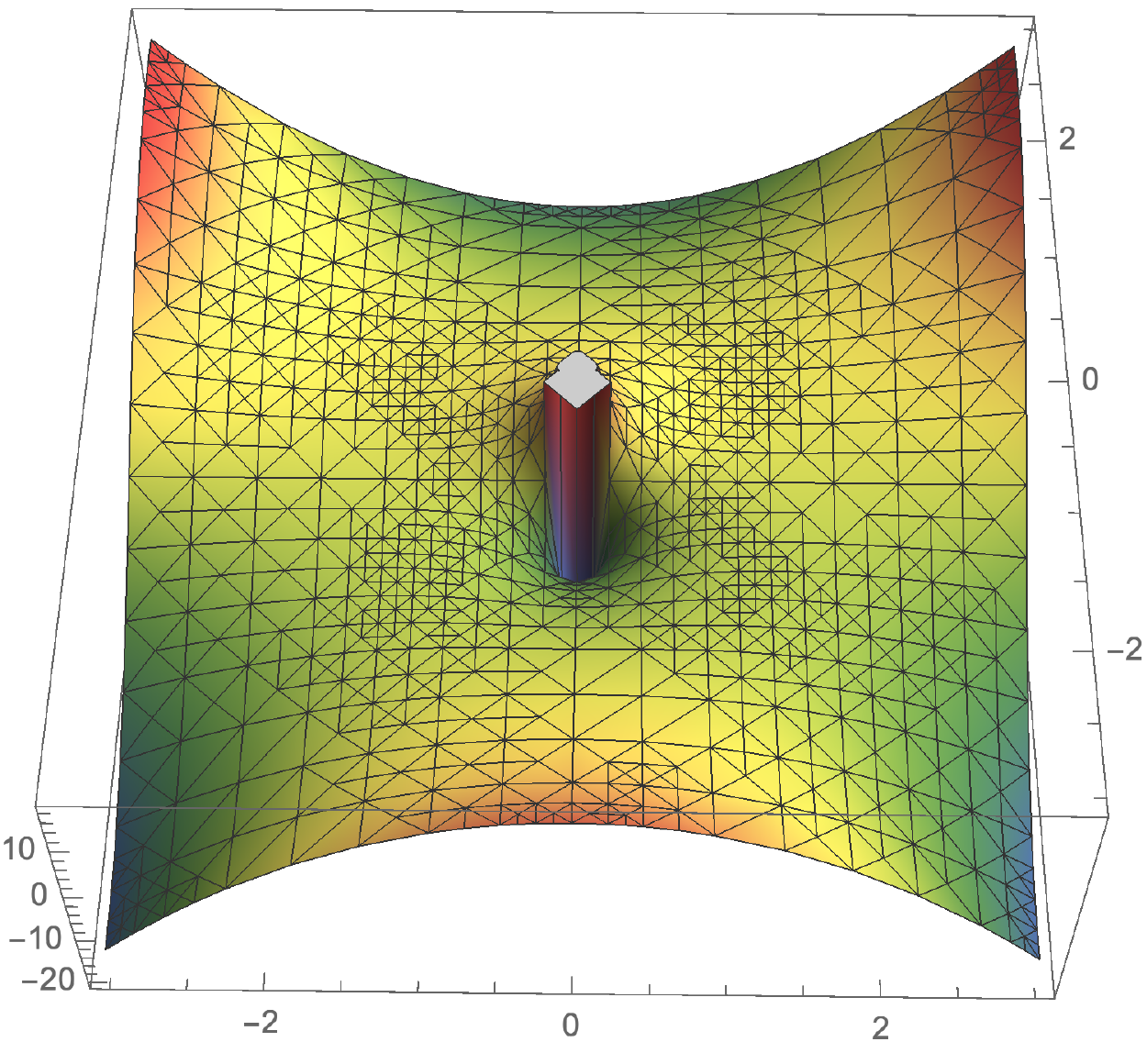}
	\includegraphics[scale=0.48]{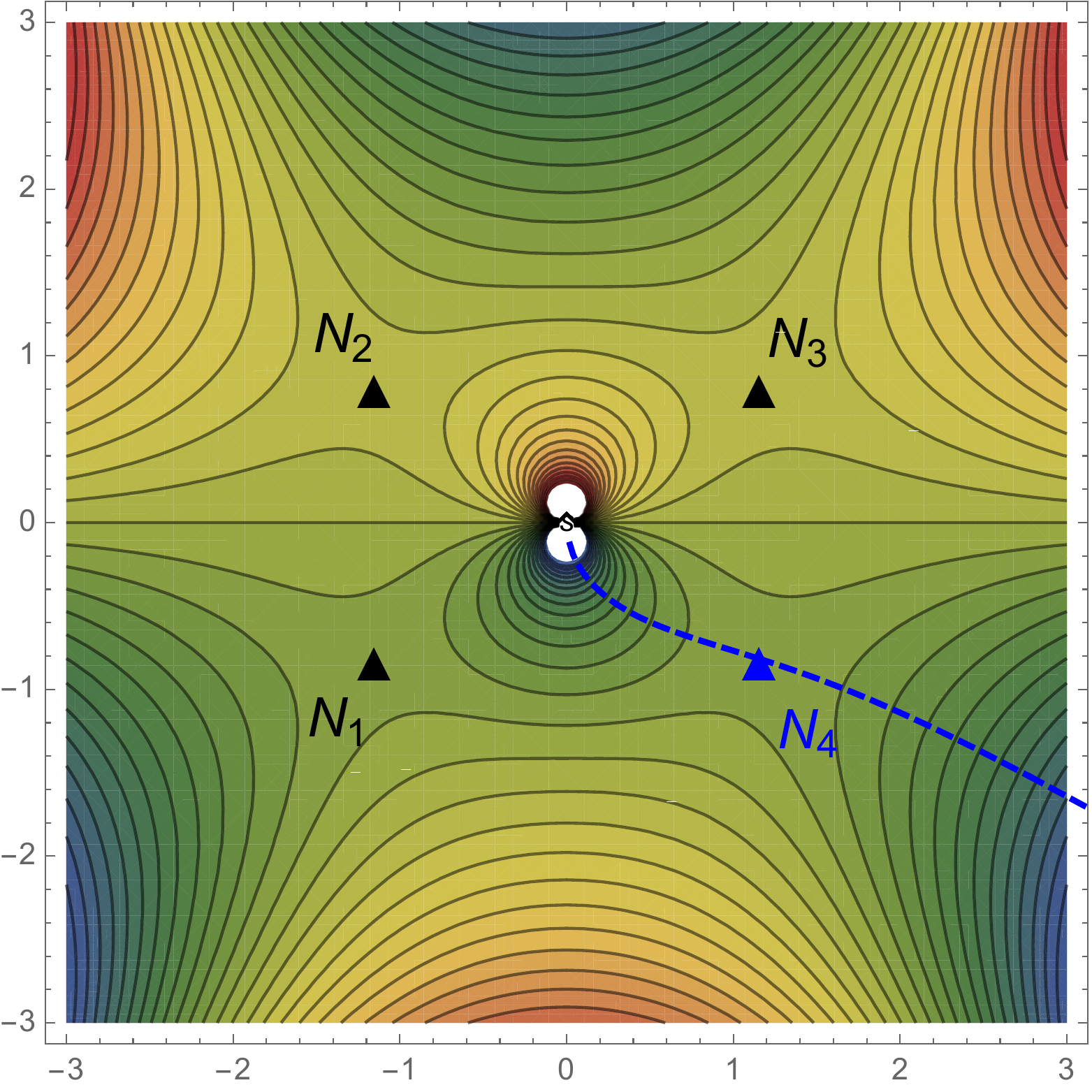}
	\includegraphics[scale=0.48]{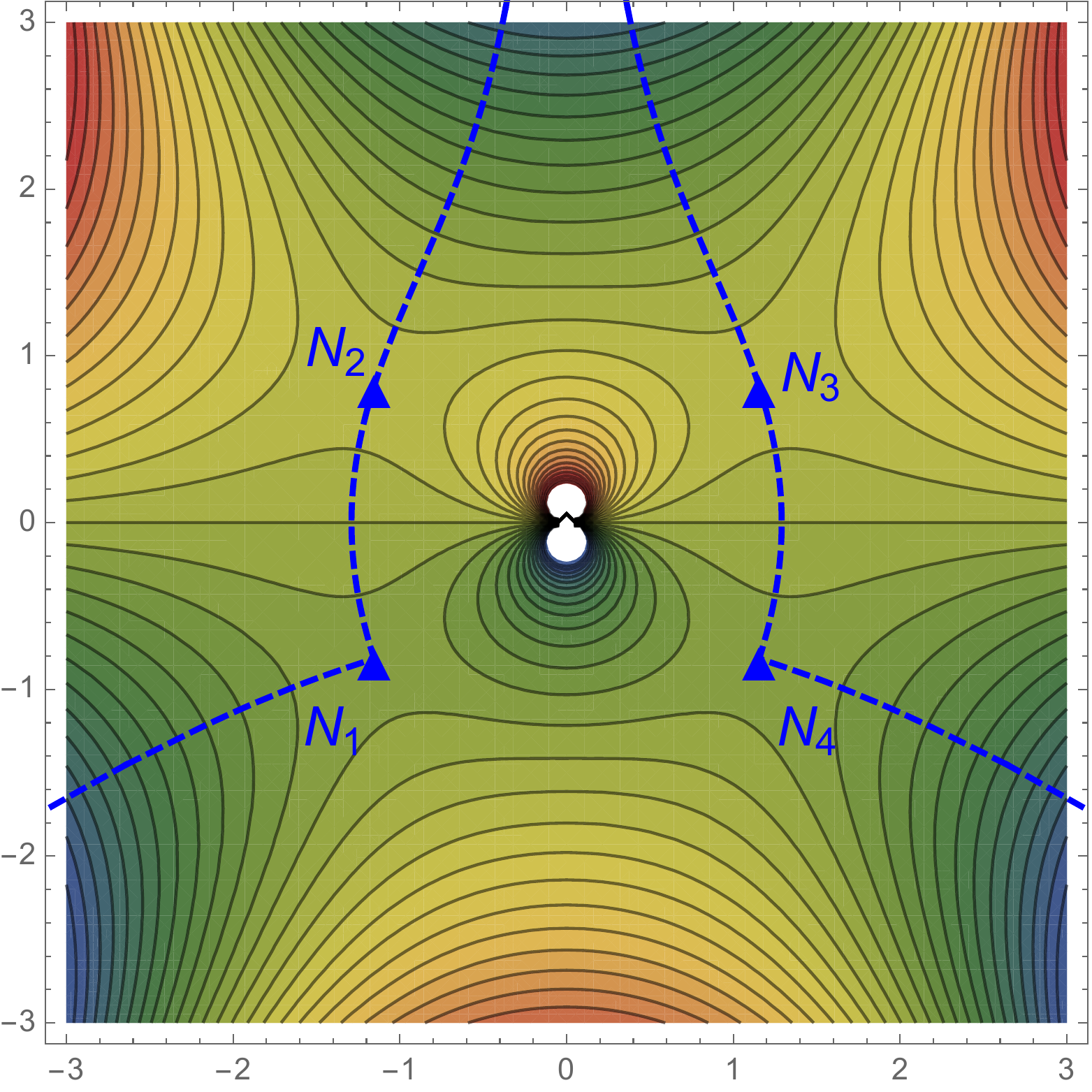}
	\includegraphics[scale=0.48]{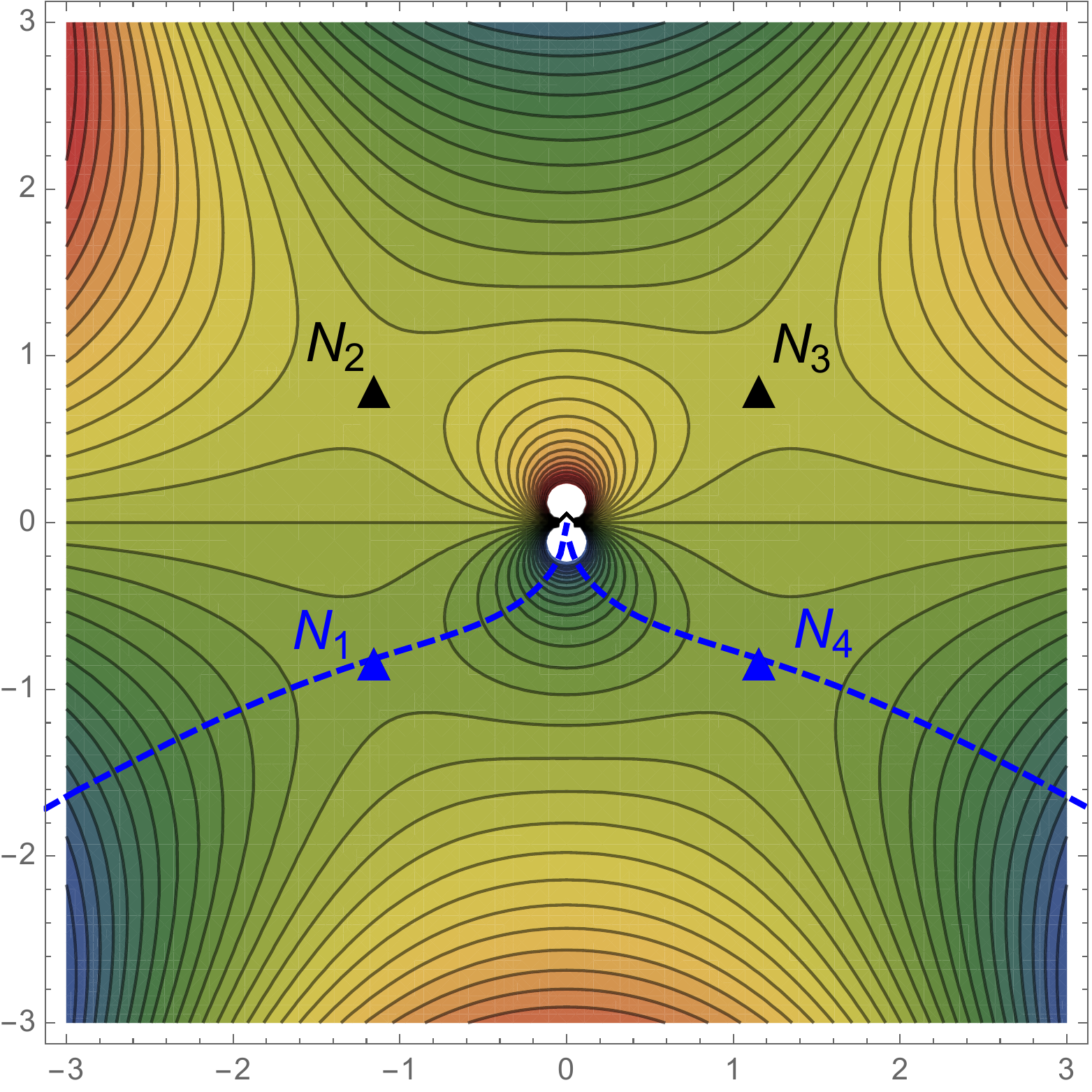}
	\caption{
These four figures show $\textrm{Re}\left[iS_0\left( N \right)\right]$ 
	in the complex plane where we set $V_0=3$, $\Lambda=3$ and $x_1=3$. 
	The $x$-axis in these figures corresponds to the real axis of the complex lapse $N$ 
	and the $y$-axis to its imaginary axis.
	The blue dashed line shows the corresponding the 
	Lefschetz thimbles with the saddle-points; $N_1= -\sqrt{\frac{2}{3}}\left( i +\sqrt{2}\right)$, $N_2= \sqrt{\frac{2}{3}}
	\left( i -\sqrt{2}\right)$,
	$N_3= \sqrt{\frac{2}{3}}\left( i +\sqrt{2}\right)$, 
	$N_4= -\sqrt{\frac{2}{3}}\left( i -\sqrt{2}\right)$.
	The upper right figure consider $N=(0,\infty)$
	and 	a Lefschetz thimble with $N_4$
       can be only chosen. The lower figures take $N=(-\infty,\infty)$ and 
       we can chose two different contours where one pass all saddle points $N_{1,2,3,4}$ and 
       another pass lower two saddle-points $N_{1,4}$. We note that 
       $N_{2,4}$ corresponds to the exponent of the WKB wave function and this point will 
       be discussed in Section~\ref{sec:Euclidean-path-integral}.}
	\label{fig:Picard-Lefschetz-linear}
\end{figure*}
%%%%%%%%%%%%%%%%%%%%%%%%%%%%%%%%%%%%%%%%%%%%%%%%
%%%%%%%%%%%%%%%%%%%%%%%%%%%%%%%%%%%%%%%%%%%%%%%%

From here we will demonstrate the Lorentzian 
Picard-Lefschetz formulation~\eqref{PL-integral} for QM.
Let us consider a simple case with
$x_0=0$, $E=0$ and $x_1 > {V_0}/{\Lambda}$.
Only one Lefschetz thimble can be chosen
in the integration domain $\mathcal{R}=(0,\infty)$.
Fig.~\ref{fig:Picard-Lefschetz-linear} discribes $\textrm{Re}\left[iS_0\left( N \right)\right]$
in the complex plane where we set $V_0=3$, $\Lambda=3$ and $x_1=3$,
where the upper right figure 
suggests that a Lefschetz thimble thorough $N_4= -\sqrt{\frac{2}{3}}\left( i -\sqrt{2}\right)$
can be only chosen in $\mathcal{R}=(0,\infty)$. Thus, 
if we consider the positive lapse $N=(0,\infty)$, we obtain the following result, 
\begin{align}\label{Lorentzian-path1}
K(x_{1}) 
& \approx \frac{e^{i\frac{\pi}{4}}}{2^{1/2}V_0^{1/4} (\Lambda x_1-V_0)^{1/4}}\\
&\times \exp \left({-\frac{2\sqrt{2}i}{3\Lambda \hbar}  \left[ \left(
-V_{0}\right)^{3/2} -\left(\Lambda x_{1}-V_{0}\right)^{3/2} \right] }\right)\notag \,.
\end{align}

For the purposes of the later discussion in Section~\ref{sec:Euclidean-path-integral}
we consider the case with $V_0={\Lambda^2}/{2}$
and $x_1 = {V_0}/{\Lambda}={\Lambda}/{2}$
and the transition amplitude is written as 
\begin{align}\label{Lorentzian-path-s1}
K(x_{1})    &\approx \exp \left({-\frac{\Lambda^2}{3\hbar}  }\right).
\end{align}
which corresponds to the Euclidean path integral
in Section~\ref{sec:Euclidean-path-integral}.

On the other hand, 
by integrating the complex lapse integral 
along $\mathcal{R}=(-\infty,\infty)$
and through the four saddle points, we obtain
\begin{align} \label{}
\begin{split}
K(x_{1})&\approx C_1\, e^{\left({-\frac{2\sqrt{2}i}{3\Lambda \hbar}  \left[ \left(
-V_{0}\right)^{3/2} +\left(\Lambda x_{1}-V_{0}\right)^{3/2} \right] }\right)}\\
&+C_2\, e^{\left({\frac{2\sqrt{2}i}{3\Lambda \hbar}  \left[ \left(
-V_{0}\right)^{3/2} -\left(\Lambda x_{1}-V_{0}\right)^{3/2} \right] }\right)}  \\
&+C_3\, e^{\left({\frac{2\sqrt{2}i}{3\Lambda \hbar}  \left[ \left(
-V_{0}\right)^{3/2} +\left(\Lambda x_{1}-V_{0}\right)^{3/2} \right] }\right)}\\
&+C_4\, e^{\left({-\frac{2\sqrt{2}i}{3\Lambda \hbar}  \left[ \left(
-V_{0}\right)^{3/2} -\left(\Lambda x_{1}-V_{0}\right)^{3/2} \right] }\right)},
\end{split} 
\end{align}
where $C$ is the prefactor at these saddle points, 
and this Lorentzian amplitude corresponds to the result of Diaz Dorronsoro 
$\textrm{et al.}$~\cite{DiazDorronsoro:2017hti}.
Strangely, therefore, all the saddle points contribute to the Lorentzian transition amplitude.
Fortheremore, 
as pointed out in Ref~\cite{Feldbrugge:2017mbc} 
choosing a different contour in $\mathcal{R}=(-\infty,\infty)$ 
leads to the different transition amplitude,
\begin{align}
\begin{split}
K(x_{1}) \approx & \ C_1\, e^{\left({-\frac{2\sqrt{2}i}{3\Lambda \hbar}  \left[ \left(
-V_{0}\right)^{3/2} +\left(\Lambda x_{1}-V_{0}\right)^{3/2} \right] }\right)}\\
&+C_4\, e^{\left({-\frac{2\sqrt{2}i}{3\Lambda \hbar}  \left[ \left(
-V_{0}\right)^{3/2} -\left(\Lambda x_{1}-V_{0}\right)^{3/2} \right] }\right)}.
\end{split} 
\end{align}
In Fig.~\ref{fig:Picard-Lefschetz-linear}
the lower figures consider $N=(-\infty,\infty)$ and 
we can chose two different contours where one pass all saddle points $N_{1,2,3,4}$ and 
 another pass lower two saddle-points $N_{1,4}$.

Let us consider the Lorentzian path integral~\eqref{QM-amplitude}
and take a different semi-classical approximation to the action~\eqref{QM-action}.
In the previous discussion, the action was semi-classically approximated by
the solution of the equation of motion~\eqref{eq:xeqn}, 
but now let us consider the semi-classical approximation of the action~\eqref{QM-action}
by the constraint equation~\eqref{eq:Neqn}.
Thus, by solving the constraint equation 
\eqref{eq:Neqn} for $\dot{x}$ and substituting in the action~\eqref{QM-action},
we get the following semi-classical action,
\begin{align}\label{sc-action}
S_0 & =  \int_{0}^{1}dt[2N(E-V)] =\int_{x_{0}}^{x_{1}}\frac{dx}{\dot{x}}2N(E-V)\nonumber \\
&= \pm\int_{x_{0}}^{x_{1}}dx\sqrt{2(E-V)},
\end{align}
where it is important to note that this semi-classical action 
is different from $S_0[N]$~\eqref{semi-action}, cancels and 
does not have the contribution of $N$~\cite{deAlwis:2018sec}.
%%%%%%%%%%%%%%%%%%%%%%%%%%
\footnote{
By using the Lorentzian path integral~\eqref{QM-amplitude}
and integrating the lapse $N$, the semi-classical path integral diverges, 
\begin{align*}
K(x_{f};x_{i})&\approx \int_\mathcal{C}\mathrm{d} N \exp\left(\frac{i S_0}{\hbar}\right) \\
&\approx e^{\pm i \int_{x_{0}}^{x_{1}}dx\sqrt{2(E-V)}/\hbar}\int^{\infty}_{0}\mathrm{d} N 
\rightarrow \infty~.
\end{align*}
}
%%%%%%%%%%%%%%%%%%%%%%%%%%

Furthermore, importantly, the saddle-point of the semi-classical action
$S_0[N]$~\eqref{semi-action} is consistent with $S_0$~\eqref{sc-action}.
In fact, in the linear potential, 
the sem-classical action is given by
\begin{align}\label{L-WKB-action}
S_{0}& =  \pm\int_{x_{0}}^{x_{1}}dx\sqrt{2(E-V)}=\pm\int_{x_{0}}^{x_{1}}dx\sqrt{2(E-V_{0}+\Lambda x)}\nonumber \\
 & = \pm\frac{2\sqrt{2}}{3\Lambda}  \left[ \left(\Lambda x_{0}+E-V_{0}\right)^{3/2} -
 \left(\Lambda x_{1}+E-V_{0}\right)^{3/2} \right],
\end{align}
which corresponds to the saddle-point action $S_0[N_s]$~\eqref{saddle-action-L}.
In Section~\ref{sec:Euclidean-path-integral}
we will discuss these coincidences in detail.
We note that the two saddle points~\eqref{saddle-lapse-L}
corresponds to the WKB approximation, but other two saddle-points are conjugate
for these saddle points.

%%%%%%%%%%%%%%%%%%%%%%%%%%%%%%%%%%%%
%%%%%%%%%%%%%%%%%%%%%%%%%%%%%%%%%%%%
\subsection{Harmonic oscillator and inverted harmonic oscillator models}
%%%%%%%%%%%%%%%%%%%%%%%%%%%%%%%%%%%%
%%%%%%%%%%%%%%%%%%%%%%%%%%%%%%%%%%%%
In the previous subsection, 
we applied the Lorentzian path integral formulation to the linear potential
and discuss some problems with the ambiguity of the lapse function.
Let us put these issues aside and 
consider this Lorentzian formulation to the harmonic oscillator 
and inverted harmonic oscillator models, which are well known in QM.

First, let us consider the harmonic oscillator model with $V=V_0+\frac{1}{2}
\Omega^2 x^2$. 
For simplicity, we consider the zero-energy system with $E=0$ and 
the solution of the equations of motion is given by
\begin{align}
x_s&=x_0\cos\left(\Omega Nt\right)-x_0\cot\left(\Omega N\right)\sin\left(\Omega Nt\right)
\notag \\
&+x_1\csc\left(\Omega N\right)\sin\left(\Omega Nt\right),
\end{align}
where we set $x(0)=x_{0}$ and $x(1)=x_{1}$.
By applying the semi-classical approximation to the 
Lorentzian path integral~\eqref{QM-amplitude}
as well as the linear potential case, we obtain the following integral,
\begin{equation}
K(x_{1};x_{0}) = \sqrt{\frac{1}{2\pi i\hbar }} \int_\mathcal{C}
\frac{\mathrm{d} N}{N^{1/2}} \exp\left(\frac{i S_0[N]}{\hbar}\right),
\end{equation}
where 
\begin{align}
\begin{split}
&S_0[N]  =\int_{0}^{1}dt N
\left( \frac{\dot{x}_s^{2}}{2N^2}-V_{0}-\frac{1}{2}
\Omega^2 x_s^2\right) \\
={}&-NV_{0}+\frac{1}{2}\left(x_0^2+x_1^2\right)\Omega\cot\left(\Omega N\right)
-x_0x_1\Omega\csc\left(\Omega N\right)\,.
\end{split}
\end{align}

%%%%%%%%%%%%%%%%%%%%%%%%%%%%%%%%%%%%%%%%%%%
%%%%%%%%%%%%%%%%%%%%%%%%%%%%%%%%%%%%%%%%%%%
\begin{figure}[t] 
	\centering
	\includegraphics[width=0.49\textwidth]{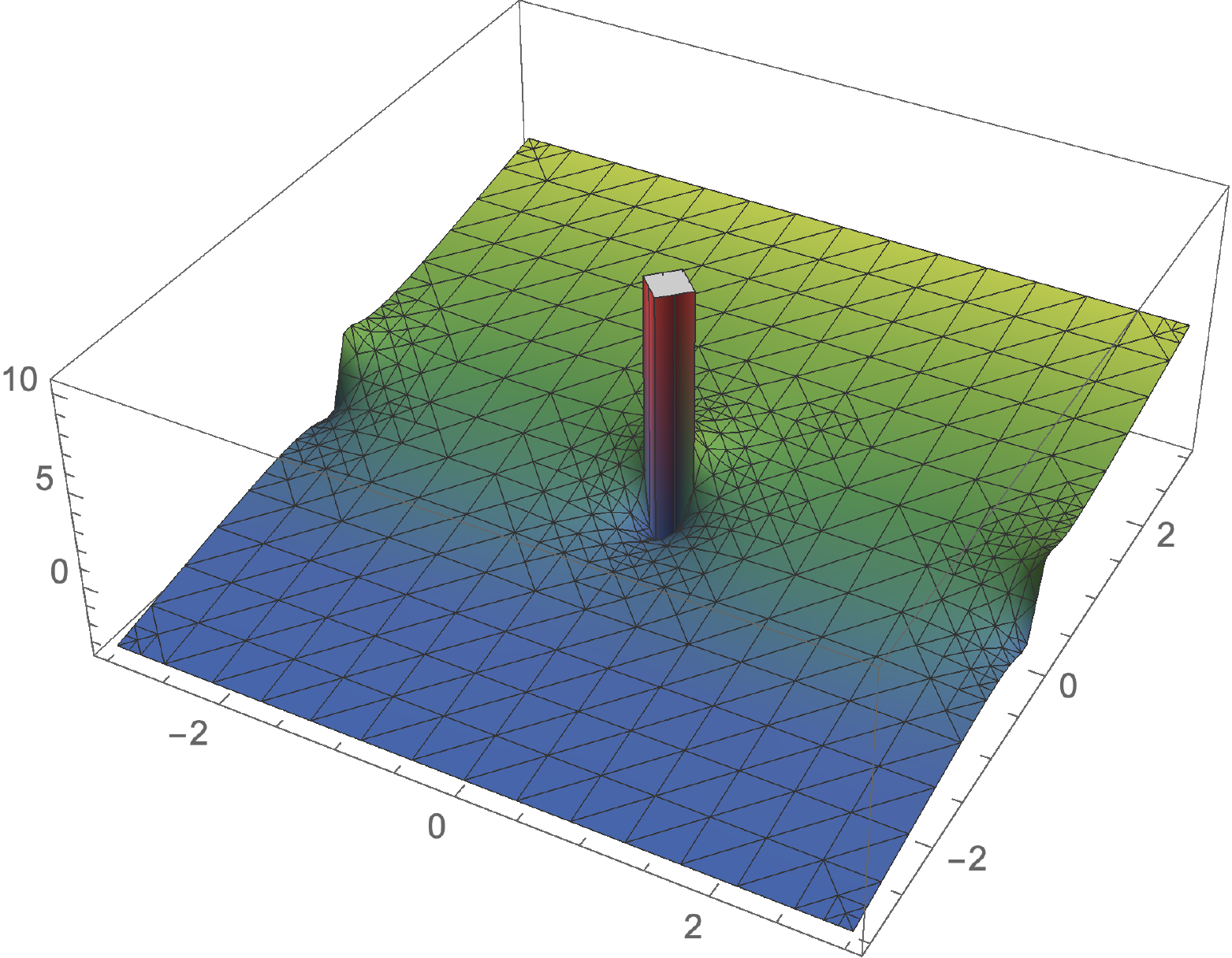}
	\includegraphics[width=0.49\textwidth]{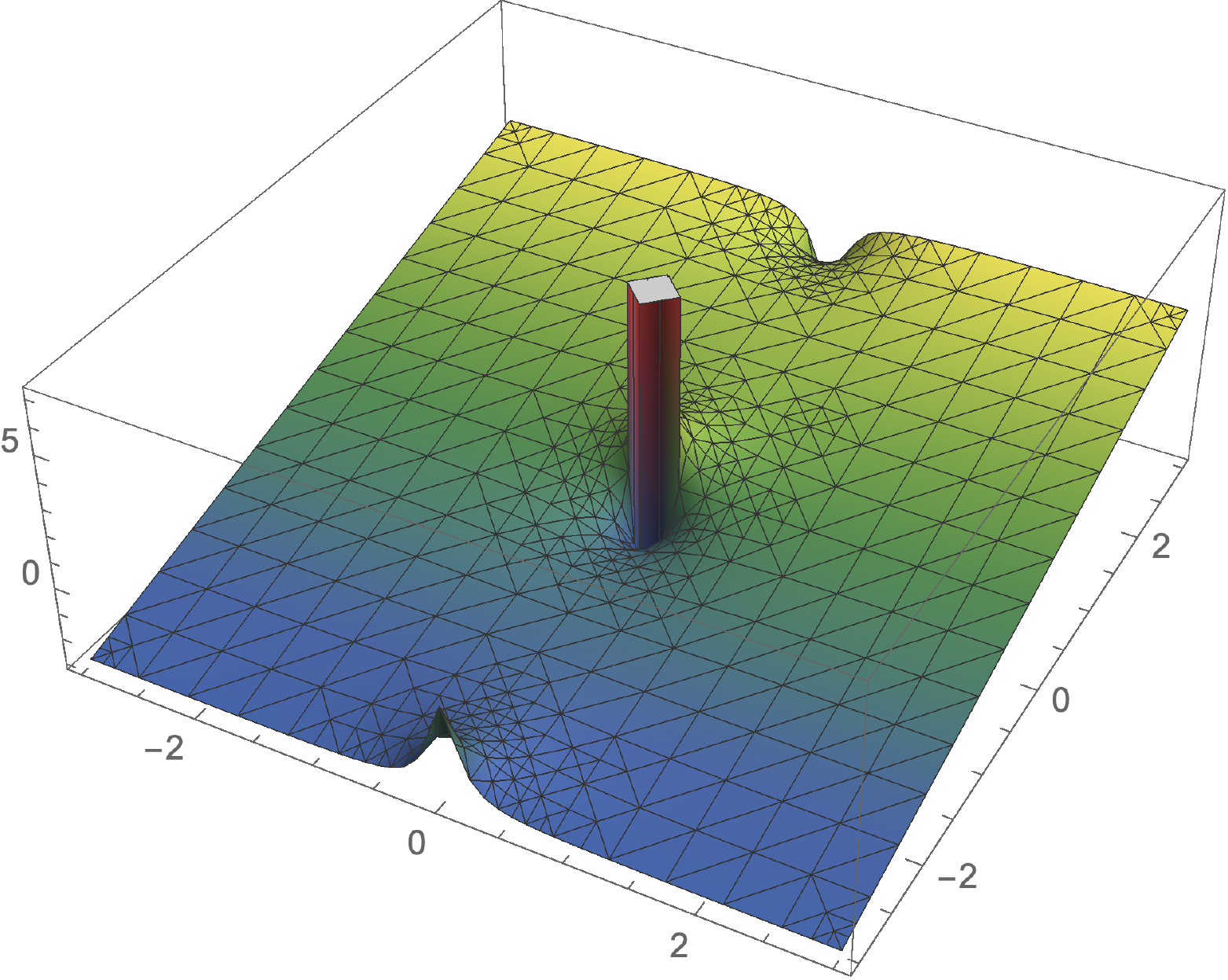}
	\caption{The top and bottom figures show 
	3D-plot of $\textrm{Re}\left[iS_0^{\, \rm saddle}\left( N \right)\right]$ 
	where we take $V_0=1$ and $x_1=1$ for 
	harmonic and inverted harmonic oscillator models, respectively.
	The Lefschetz thimbles ${\cal J}_{\sigma}$ on the lapse
	integral is taken along the imaginary 
	$y$-axis in the harmonic oscillator whereas it is taken along the real $x$-axis
	for the inverse harmonic oscillator.}
	\label{fig:Picard-Lefschetz-harmonic}
\end{figure} 
%%%%%%%%%%%%%%%%%%%%%%%%%%%%%%%%%%%%%%%%%%%
%%%%%%%%%%%%%%%%%%%%%%%%%%%%%%%%%%%%%%%%%%%

When we take $x_0=0$ and $\Omega=1$, 
the semi-classical action $S_0[N]$ reads,
\begin{align}
S_0[N] =-NV_{0}+\frac{1}{2}x_1^2\cot\left(N\right)\,.
\end{align}
By solving ${\partial S_0[N]}/{\partial N}=0$,
the saddle-points are given as,
\begin{align}
&\sin^2\left({N_s}\right)=-\frac{x_1^2}{2V_{0}}\ \Longleftrightarrow \ \\
&N_s=\pm i\sinh^{-1}\sqrt{\frac{x_1^2}{2V_{0}} }+2\pi c_1,\ 
\pi \pm i\sinh^{-1}\sqrt{\frac{x_1^2}{2V_{0}} }+2\pi c_1\,,\notag 
\end{align}
where $c_1\in\mathbb{Z}$. 
In Fig.~\ref{fig:Picard-Lefschetz-harmonic}
 we show the contour plot of
$\textrm{Re}\left[iS_0^{\, \rm saddle}\left( N \right)\right]$
in the complex plane where we set $V_0=1$ and $x_1=1$ for
the harmonic and inverted harmonic oscillator models.
Although there are many saddle points, 
we will simply 
consider the contour corresponding to $c_1=0$ in the integration domain $\mathcal{R}=(0,\infty)$.

Thus, we take one saddle-point
$N_s=- i\sinh^{-1}\sqrt{\frac{x_1^2}{2V_{0}} }$
and the saddle-point action is evaluated as 
\begin{align}\label{saddle-action-H}
S_0[N_s] &= -N_sV_{0}+\frac{1}{2}x_1^2\cot\left(N_s\right)\nonumber\\
&= iV_{0}\,\sinh^{-1}\sqrt{\frac{x_1^2}{2V_{0}} }
+\frac{ix_1\sqrt{V_0}}{2}\sqrt{2+\frac{x_1^2}{V_{0}} }\,,
\end{align}
and we obtain the following transition amplitude,
\begin{align}\label{Lorentzian-path-h}
K(x_{1})  &\approx \exp \left[\frac{-1}{\hbar}\left( V_{0}\,\sinh^{-1}\sqrt{\frac{x_1^2}{2V_{0}} }
+\frac{x_1\sqrt{V_0}}{2}\sqrt{2+\frac{x_1^2}{V_{0}} }\right)\right].
 \end{align}
For the purposes of the later discussion in Section~\ref{sec:Euclidean-path-integral} 
let us consider the specific case which satisfy 
\begin{align}
x_1= \frac{\left(e^2-1\right) \sqrt{V_0}}{\sqrt{2} e},
\end{align}
and the transition amplitude reads
\begin{align}\label{Lorentzian-path-h1}
K(x_1)    &\approx \exp \left[ \frac{ \left(1-4 e^2-e^4\right) V_0}{4\hbar\, e^2}\right].
\end{align}

Next, let us consider the inverted harmonic oscillator model with $V=V_0
-\frac{1}{2}
\Omega^2 x^2$. 
For simplicity, we consider the zero-energy system with $E=0$ and 
the solution of the system is given by
\begin{equation}
x_s=\frac{e^{-\Omega N t } \left(-x_0 e^{2 \Omega N t }+x_1 e^{2 \Omega N t +\Omega N }+
x_0 e^{2 \Omega N }-x_1 e^{\Omega N}\right)}{e^{2 \Omega N}-1}.
\end{equation}
By taking semi-classical approximation to the 
Lorentzian path integral~\eqref{QM-amplitude}
we get the following semi-classical action,
\begin{align}
\begin{split}
&S_0[N]  =\int_{0}^{1}dt N
\left( \frac{\dot{x}_s^{2}}{2N^2}-V_{0}-\frac{1}{2}
\Omega^2 x_s^2\right) \\
=&-N V_{0}+\frac{1}{2} \Omega  \left(x_0^2+x_1^2\right) \coth (N \Omega )-x_0 x_1\Omega
\, \text{csch}(N \Omega )\,.
\end{split}
\end{align}
We set $x_0=0$ and $\Omega=1$, and $S_0[N]$ reads,
\begin{align}
S_0[N] =-NV_{0}+\frac{1}{2}x_1^2\coth\left(N\right)\,.
\end{align}
By solving ${\partial S_0[N]}/{\partial N}=0$,
the corresponding saddle-points are given by,
\begin{align}
\begin{split}
&\sinh^2\left({N_s}\right)=-\frac{x_1^2}{2V_{0}}\ \Longleftrightarrow \ \\
N_s=\pm i\sin^{-1}&\sqrt{\frac{x_1^2}{2V_{0}} }+2i\pi c_2,\ 
i\pi \pm i\sin^{-1}\sqrt{\frac{x_1^2}{2V_{0}} }+2i\pi c_2\,,
\end{split}
\end{align}
where $c_2\in\mathbb{Z}$. 
As before, we consider the contour corresponding to $c_2=0$ 
in $\mathcal{R}=(0,\infty)$. 
Hence, we take $N_s=- i\sin^{-1}\sqrt{\frac{x_1^2}{2V_{0}} }$
and the saddle-point action $S_0[N_s]$ is evaluated as 
\begin{align}\label{saddle-action-I}
S_0[N_s]&= -N_sV_{0}+\frac{1}{2}x_1^2\coth\left(N_s\right)\nonumber\\
&= iV_{0}\,\sin^{-1}\sqrt{\frac{x_1^2}{2V_{0}} }
+\frac{ix_1\sqrt{V_0}}{2}\sqrt{2-\frac{x_1^2}{V_{0}} }\,.
\end{align}
For the later discussion in Section~\ref{sec:Euclidean-path-integral}, 
let us consider the following case,
\begin{align}
x_1= \sqrt{2V_0} \sin (1),
\end{align}
and the transition amplitude reads
\begin{align}\label{Lorentzian-path-i1}
K(x_1)    &\approx \exp \left(- \frac{ V_0+V_0 \sin (1) \cos (1)}{\hbar}\right).
\end{align}

%%%%%%%%%%%%%%%%%%%%%%%%%%%%%%%%%%%%
%%%%%%%%%%%%%%%%%%%%%%%%%%%%%%%%%%%%
\section{Euclidean path integral, instanton and WKB approximation}
\label{sec:Euclidean-path-integral}
%%%%%%%%%%%%%%%%%%%%%%%%%%%%%%%%%%%%
%%%%%%%%%%%%%%%%%%%%%%%%%%%%%%%%%%%%
In this section, we review the Euclidean path integral, 
instanton and WKB approximation for 
the quantum tunneling and discuss their relations to the 
Lorentzian Picard-Lefschetz formulation~\eqref{PL-integral}.

%%%%%%%%%%%%%%%%%%%%%%%%%%%%%%%%%%%%
%%%%%%%%%%%%%%%%%%%%%%%%%%%%%%%%%%%%
\subsection{Euclidean path integral and instanton}
%%%%%%%%%%%%%%%%%%%%%%%%%%%%%%%%%%%%
%%%%%%%%%%%%%%%%%%%%%%%%%%%%%%%%%%%%
The evolution of the wave function in the classical regime can 
be approximated by saddle points $\delta S[x]/\delta x \approx 0$
which satisfy the classical equation of motion. 
On the other hand, for quantum tunneling path which is 
the classically forbidden region 
the transition amplitude is 
approximately given by the saddle points of the 
Euclidean path integral 
which is derived by the solution of the Euclidean equations of motion.

We can easily show that setting $N=-i$
the Lorentzian path integral~\eqref{QM-amplitude} 
corresponds to the Euclidean amplitude.
In fact, the reparameterized action $S[x,N]$ is given by $t\rightarrow Nt$
in $S[x]$ and the Euclidean action $S_E[x]$ 
is given by $t\rightarrow -i\tau$.
Therefore, the action $S[x,-i]$ represents the 
Euclidean action $S_E[x]$,
\begin{align}\label{Euclidean-action}
&iS[x,-i]=-\int_{t_i}^{t_f}dt
\left(\frac{\dot{x}^{2}}{2}+V(x)\right) \\ & \iff
iS[x] \equiv -S_E[x]= -\int_{\tau_i}^{\tau_f} \mathrm{d}\tau 
\left(\frac{1}{2}\left(\frac{d x}{d\tau}\right)^2 + V(x) \right)\,,\notag
\end{align}
where the potential changes sign $V(x)\rightarrow -V(x)$
in $S_E[x]$.
From \eqref{Euclidean-action} we derive the following Euclidean equations of motion,
\begin{align}
\frac{d^2 x}{d\tau^2}=V'(x).
\end{align}

However, the extrema of $N$ corresponding to the saddle points of the Lorentzian path integral~\eqref{QM-amplitude} using the Picard-Lefschetz theory deviate from $N = -i$ and do not reproduce the transition amplitude
based on the Euclidean path integral. 
From here we show some examples to clarify this fact.
Let us consider the linear potential $V=V_{0}-\Lambda x$ with $\Lambda>0$
for simplicity.
Note that solving the Euclidean equations of motion and 
substituting the solutions for the Euclidean 
action $S_E[x]$ leads to the instanton amplitude. 
Therefore, the oscillatory integral~\eqref{o-integral}
is consistent with the Euclidean amplitude except for 
the lapse integral. Thus, by taking $N=-i$
we have the Euclidean transition amplitude for the linear potential,
\begin{equation}\label{instanton}
K(x_{1};x_{0}) = \sqrt{\frac{-1}{2\pi \hbar }}  e^{iS_0[-i]/\hbar}
= \sqrt{\frac{-1}{2\pi \hbar }}  e^{-S_E/\hbar},
\end{equation}
where $S_E$ is 
\begin{align}
\begin{split}
S_E&= \int_{\tau_i=0}^{\tau_f=1} \mathrm{d}\tau 
\left\{\frac{1}{2}\left(\frac{d x_s}{d\tau}\right)^2 +V_{0}-\Lambda x_s \right\} \\
&= -\frac{\Lambda^{2}}{24}-\frac{\Lambda}{2}(x_{1}+x_{0})+V_{0}+\frac{(x_{1}-x_{0})^{2}}{2}\,.
\end{split}
\end{align}
For simplicity, we consider
$x_0=0$ and the transition amplitude is given by
\begin{align} 
K(x_{1})    &\approx \exp \left[{-\frac{1}{\hbar}  \left(-\frac{\Lambda^{2}}{24}-\frac{\Lambda x_{1}}{2}+V_{0}+\frac{x_{1}^{2}}{2}\right) }\right],\end{align}
which is not consistent with the result~\eqref{Lorentzian-path1} 
of the Lorentizan path integral.
As we will see later, this result is not compatible with the WKB approximation either.

\begin{figure}[t]
	\centering
	\includegraphics[scale=0.6]{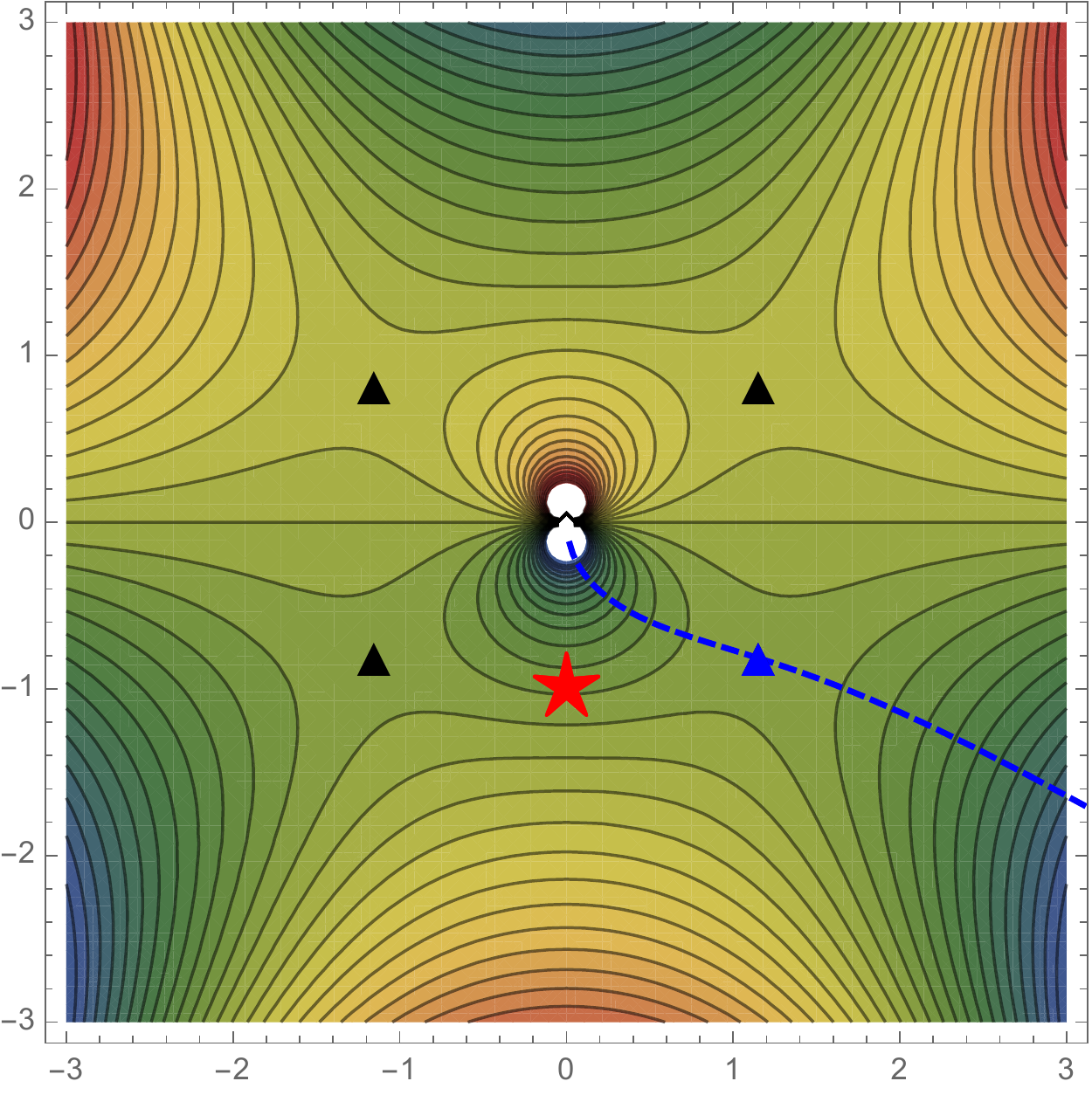}
	\includegraphics[scale=0.6]{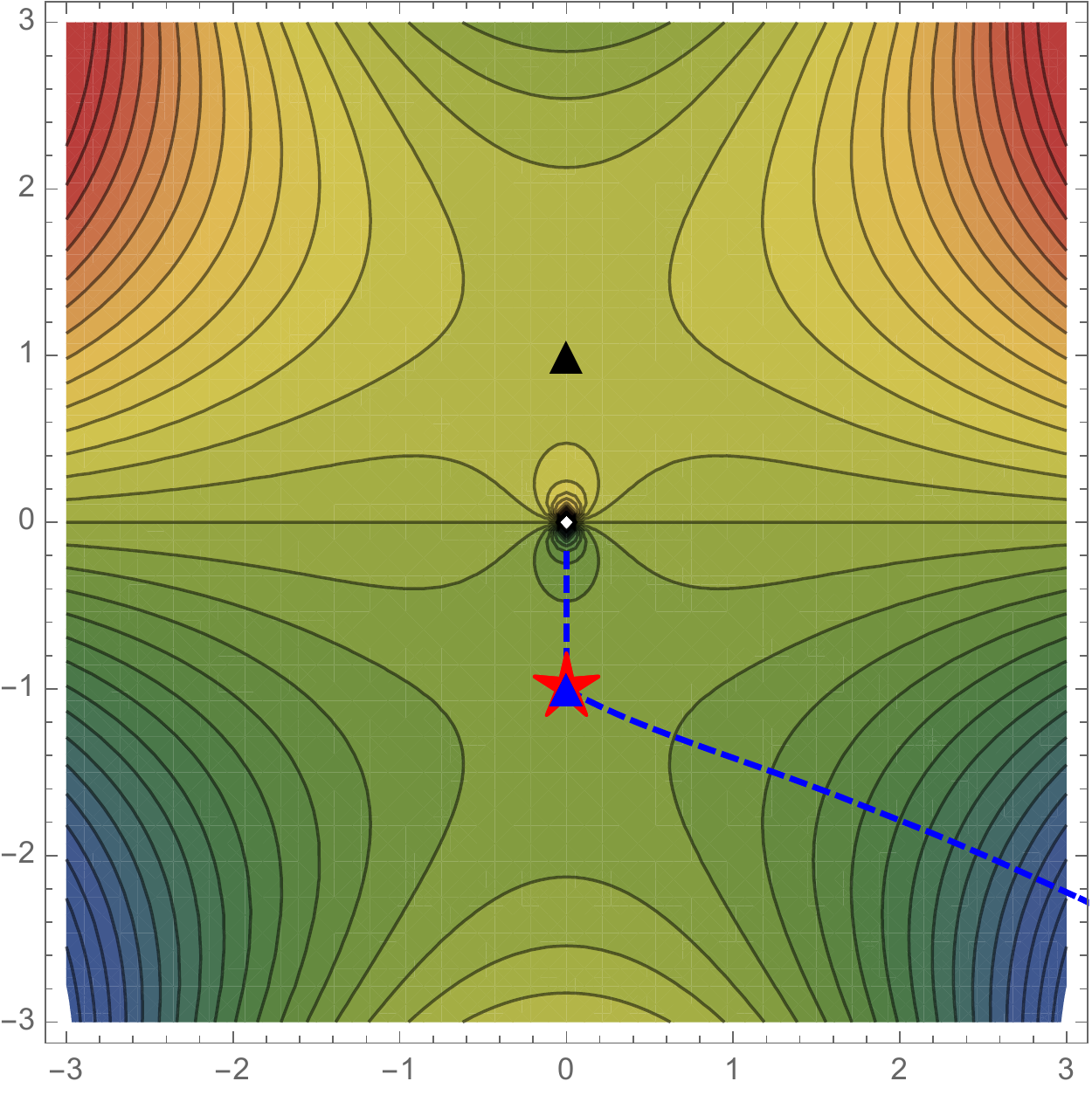}
	\caption{
The top figure shows $\textrm{Re}\left[iS_0^{\, \rm saddle}\left( N \right)\right]$ in complex plane where we set $V_0=3$, $\Lambda=3$ and $x_1=3$
and the star expresses $N=-i$.
On the other hand,  in the bottom figure
we set $V_0=9/2$, $\Lambda=3$ and $x_1=3/2$
and this figure shows that the saddle points of 
the action~\eqref{saddle-action-L} is $N_s$
also coincides with the Euclidean saddle point $N=-i$.}
\label{fig:Picard-Lefschetz-Euclidean}
\end{figure}

The reason is that the semi-classical solution~\eqref{x-semisol} with 
$N=-i$ does not satisfy the constraint equation~\eqref{eq:Neqn}, 
which is the law of conservation of energy. 
Thus, when the constraint equation~\eqref{eq:Neqn} is actually satisfied
the result of the Euclidean path integral coincides with 
the Lorentzian Picard-Lefschetz formulation~\eqref{Lorentzian-path1}.
For instance, when we take $V_0={\Lambda^2}/{2}$
and $x_1 = {V_0}/{\Lambda}={\Lambda}/{2}$
and the transition amplitude is given by
\begin{align}
K(x_{1})    &\approx \exp \left({-\frac{\Lambda^2}{3\hbar}  }\right),
\end{align}
which is consistent with the result~\eqref{Lorentzian-path-s1} 
in the Lorentzian Picard-Lefschetz formulation. 
In this case the saddle points of 
the action~\eqref{saddle-action-L}, which is $N_s=\pm i$
also coincides with the Euclidean saddle point $N=-i$.
In Fig.~\ref{fig:Picard-Lefschetz-Euclidean} 
we show this correspondence.

%%%%%%%%%%%%%%%%%%%%%%%%%%%%%%%%%%%%%%%%%%%%%%%%
%%%%%%%%%%%%%%%%%%%%%%%%%%%%%%%%%%%%%%%%%%%%%%%%
%%%%%%%%%%%%%%%%%%%%%%%%%%%%%%%%%%%%%%%%%%%%%%%%
%%%%%%%%%%%%%%%%%%%%%%%%%%%%%%%%%%%%%%%%%%%%%%%%

Next, let us discuss the well-known double well potential, 
\begin{equation}
V(x)=\frac{\lambda}{4}\left(x^2-a^2 \right)^2,
\end{equation}
and consider the quantum tunneling from $x_0=-a$ to $x_1=a$.
The usual method for finding the Euclidean (instanton) solution 
in this case, is to obtain it directly from the Euclidean energy conservation 
rather than solving the Euclidean equations of motion.
For simplicity, we set $E=0$ and the Euclidean energy conservation gives,
\begin{equation}
\frac{1}{2}\left(\frac{d x}{d\tau}\right)^2-V(x)=0\ \Longrightarrow \ 
\frac{d x}{d\tau}=\pm\sqrt{\frac{\lambda}{2}}\left(x^2-a^2 \right).
\end{equation}
Thus, we can get the following instanton solution interpolating between $-a$ and $a$,
\begin{equation}
x(\tau)=\pm \,a\tanh\frac{\omega}{2}\left(\tau-\tau_0\right),
\end{equation}
where $\omega=\sqrt{2\lambda a^3}$
and $\tau_0$ is an integration constant.
The plus and minus classical solutions are the instanton and 
anti-instanton.
The instanton corresponds to the particle initially 
sitting on the maximum of $-V(x)$ at $x=-a$, 
passing $x=0$  for a very short time and 
ending up at the other maximum of $-V(x)$ at $x=a$.
From the instanton
the Euclidean action is given by
\begin{align} \label{instanton-action}
S_E=\frac{2 \sqrt{2\lambda}\, a^3  }{3 },
\end{align}
whose the transition amplitude $K(a)$ is consistent with 
the WKB approximation of the wave function.
For instance, if we extend this instanton as follows,
\begin{equation}
x(t)=a\tanh\frac{\omega}{2}\left(iNt -\tau_0\right),
\end{equation}
and apply it to the Lorentzian Picard-Lefschetz formulation, it returns the same result. As already discussed, a saddle-point approximation of the action by a solution of the Euclidean equations of motion does not give the correct transition amplitude. The correct result is obtained when the solution of the Euclidean equation of motion satisfies the energy conservation. Since the instanton solution is derived from the energy conservation in Euclidean form, it necessarily corresponds to the saddle point of the Lorentzian path integral.
It is important to note that the solutions of the Euclidean equation of motion do not necessarily correspond to the correct semiclassical saddle point solutions. It is physically meaningful only if the solutions satisfy the constraint equation~\eqref{eq:Neqn}.

We can see the same relations with the harmonic oscillator 
and inverted harmonic oscillator models.
By taking $N=-i$
we have the Euclidean transition amplitude for the harmonic 
and inverted harmonic oscillator,
\begin{align} 
K^H(x_{1};x_0)    &\approx e^{\left[{-\frac{1}{\hbar}  \left(V_{0}+\frac{1}{2}\left(x_0^2+x_1^2\right)\Omega\coth\Omega
-x_0x_1\Omega\,\text{csch}\,\Omega\right) }\right]}, \\
K^I(x_{1};x_0)&\approx e^{\left[{-\frac{1}{\hbar}  \left(V_{0}+\frac{1}{2}\left(x_0^2+x_1^2\right)\Omega\cot\Omega
-x_0x_1\Omega\,\text{csc}\,\Omega\right) }\right]},
\end{align}
where we denote that $K^H(x_{1};x_0) $, $K^I(x_{1};x_0)$
are the transition amplitudes for the harmonic
and inverted harmonic oscillator
are not consistent with the Lorentizan 
formulations~\eqref{Lorentzian-path-h}. 
As discussed previously, when we take $x_0=0$ and $\Omega=1$
and choose $x_1$ which satisfies 
the Euclidean energy conservation,
\begin{align}
x_1^H= \frac{\left(e^2-1\right) \sqrt{V_0}}{\sqrt{2} e},\quad 
x_1^I=   \sqrt{2V_0} \sin (1),
\end{align}
the Euclidean transition amplitudes
are consistent with 
the results~\eqref{Lorentzian-path-h1} and~\eqref{Lorentzian-path-i1}
of the Lorentzian formulation,
\begin{align} 
K^H(x_{1})    &\approx \exp \left(\frac{ \left(1-4 e^2-e^4\right) V_0}{4\hbar e^2}\right), \\
K^I(x_{1})&\approx \exp \left(-\frac{ V_0+V_0 \sin (1) \cos (1)}{\hbar}\right).
\end{align}

\medskip
%%%%%%%%%%%%%%%%%%%%%%%%%%%%%%%%%%%%
%%%%%%%%%%%%%%%%%%%%%%%%%%%%%%%%%%%%
\subsection{WKB approximation of Schr\"{o}dinger equation}
%%%%%%%%%%%%%%%%%%%%%%%%%%%%%%%%%%%%
%%%%%%%%%%%%%%%%%%%%%%%%%%%%%%%%%%%%
Let us discuss the WKB approximation 
of the Schr\"{o}dinger equation and the correspondence 
to the Lorentzian and Euclidean path integral formulation.
The WKB approximation (or WKB method) is one of the semi-classical 
approximation methods for the Schr\"{o}dinger equation. 
For the Schr\"{o}dinger equation, 
which is the fundamental equation of QM and reads, 
\begin{equation}\label{Schrodinger-eq}
\hat{H}\Psi(x)=\left(-\frac{\hbar^{2}}{2}\frac{d^{2}}{dx^{2}}+V(x)\right)\Psi(x)=E\, \Psi(x),
\end{equation}
where $\hat{H}$ is the Hamiltonian,
we assume that the solution is in the form of $\exp\, (\frac{i}{\hbar}S[x])$ 
and expanded as a perturbation series of $\hbar$.
By substituting $\Psi(x)\approx 
e^{\frac{i}{\hbar}\left(S_{0}[x]+\hbar S_{1}[x]+\cdots\right)}$
in the Schr\"{o}dinger equation~\eqref{Schrodinger-eq}
we obtain the following equations,
\begin{equation}
\frac{1}{2}\left(\frac{dS_{0}}{dx}\right)^{2}+V-E=0,\quad 
\frac{dS_{1}}{dx}=\frac{i}{2}\frac{d}{dx}\left(\ln\frac{dS_{0}}{dx}\right)\dots,
\end{equation}
where $S_{0}$ is the dominant contribution of the WKB wave function
and can also be obtained by the constraint equation 
\eqref{eq:Neqn} as already discussed in Section~\ref{sec:Lorentzian-path-integral}.
We note that the Schr\"{o}dinger equation~\eqref{Schrodinger-eq} 
and wave function do not have the contribution 
of $N$ even if the semi-classical action includes the lapse function $N$.

In the leading order of 
the WKB approximation the wave function is given by
\begin{equation}
\Psi(x)\approx \frac{c}{|2(E-V(x))|^{1/4}}
e^{\left[{\pm\frac{i}{\hbar}\int^{x}_{x_0}dx\sqrt{2(E-V(x))}}\right]}\,.
\end{equation}
For the linear potential the WKB wave function is given by
 \begin{align}
\begin{split}
&\Psi(x) \approx  \ \frac{c_1e^{ \left({+\frac{2\sqrt{2}i}{3\Lambda \hbar}  \left[ \left(\Lambda x_{0}+E-V_{0}\right)^{3/2} -\left(\Lambda x_{1}+E-V_{0}\right)^{3/2} \right] }\right)}}{|2(E-V(x))|^{1/4}}\, \\
&+\frac{c_2e^{ \left({-\frac{2\sqrt{2}i}{3\Lambda \hbar}  \left[ \left(
\Lambda x_{0}+E-V_{0}\right)^{3/2} -\left(\Lambda x_{1}+E-V_{0}\right)^{3/2} \right] }\right)}}{|2(E-V(x))|^{1/4}}\, ,
\end{split} 
\end{align}
where the exponent of the above WKB wave function
agrees with the two saddle-point actions $S_0[N_s]$~\eqref{saddle-action-L} of the Lorentzian path integral.
In the Lorentzian 
Picard-Lefschetz formulation~\eqref{PL-integral}, four-saddle points dominate the path integral, 
and only one saddle point contributes when the lapse integral is defined as positive. 
On the other hand, in the WKB analysis, there is no uncertainty of the lapse function, 
and the exponents of the wave function can be either positive or negative, depending on the initial conditions. Thus, the positive and negative lapse function in the Lorentzian 
Picard-Lefschetz formulation~\eqref{PL-integral} could be considered.

On the other hand, for the harmonic and inverted harmonic oscillator potentials
where we take $x_0=0$ and $E=0$, the exponents of the WKB wave function
are given by
\begin{align}
S_{0}& =  \pm iV_{0}\,\sinh^{-1}\sqrt{\frac{x_1^2}{2V_{0}} }
\pm\frac{ix_1\sqrt{V_0}}{2}\sqrt{2+\frac{x_1^2}{V_{0}} }\,,\label{H-WKB-action}\\
S_{0}&= \pm iV_{0}\,\sin^{-1}\sqrt{\frac{x_1^2}{2V_{0}} }
\pm\frac{ix_1\sqrt{V_0}}{2}\sqrt{2-\frac{x_1^2}{V_{0}} }\,,\label{I-WKB-action}
\end{align}
which are exactly consistent with the saddle points of the 
Lorentzian formulation~\eqref{saddle-action-H} and \eqref{saddle-action-I}.
Finally, we comment the double well potential~\eqref{QM-action} 
with zero-energy system $E=0$ and the corresponding
semi-classical action is given by
\begin{align}
S_{0}& =  \pm\int_{x_{0}=-a}^{x_{1}=a}dx\sqrt{-\frac{\lambda}{2}\left(x^2-a^2 \right)^2}=\pm
i\frac{2 \sqrt{2\lambda}\, a^3  }{3 },
\end{align}
which is consistent with the Euclidean action $S_E$
utilizing the instanton~\eqref{instanton-action}.
In summary 
the Lorentzian Picard-Lefschetz formulation~\eqref{PL-integral} 
including the lapse $N$ integral and using the Picard-Lefschetz theory, and 
the Euclidean formulation utilizing instanton
are nothing more than the WKB analysis of the Schr\"{o}dinger equation.

The reason why these approaches 
correspond is as follows: Applying the method of steepest descents or saddle-point using the Picard-Lefschetz theory to the integration of the lapse 
$N$ corresponds to taking semi-classical contours such that 
the constraint equation~\eqref{eq:Neqn} is satisfied. 
Therefore, the Lorentzian Picard-Lefschetz formulation~\eqref{PL-integral} 
corresponds to the WKB approximation to the wave function whose 
$S_{0}$ is given by the constraint equation~\eqref{eq:Neqn}.
Conversely, in order for the Euclidean path integral $S_E$ to be the correct semiclassical 
approximation, the solutions of the Euclidean equation of motion must satisfy the constraint equation~\eqref{eq:Neqn}.

\medskip
%%%%%%%%%%%%%%%%%%%%%%%%%%%%%%%%%%%%
%%%%%%%%%%%%%%%%%%%%%%%%%%%%%%%%%%%%
\section{Lorentzian Picard-Lefschetz Formulation for quantum gravity}
\label{sec:Lorentzian-gravity} 
%%%%%%%%%%%%%%%%%%%%%%%%%%%%%%%%%%%%
%%%%%%%%%%%%%%%%%%%%%%%%%%%%%%%%%%%%
In this section we demonstrate that 
the tunneling or no-boundary wave functions derived by 
the Lorentzian Picard-Lefschetz Formulation corresponds to the WKB solution 
of the Wheeler-DeWitt equation~\cite{Brown:1990iv,Vilenkin:2018dch,deAlwis:2018sec}.

Following~\cite{Halliwell:1988ik} the gravitational amplitude 
based on Arnowitt, Deser and Misner (ADM) formalism~\cite{Arnowitt:1962hi}
can be written by the Lorentzian path integral,
\begin{equation}\label{G-amplitude2}
G(q_{f};q_{i})= \int_\mathcal{C}\mathrm{d} N \int \mathcal{D}q(t)~ 
\exp \left(\frac{i S[N,q]}{\hbar}\right)~.
\end{equation}
where $S[N,q]$ is the gravitational action with $q=a^2$.
Since $S[N,q]$ is quadratic,
the functional integral~\eqref{G-amplitude2} can be evaluated 
under the semi-classical approximation.
We assume $q(t) = q_s(t) + Q(t)$ where
$Q(t)$ is the Gaussian fluctuation around the semi-classical solution.
By substituting it for $S[N,q]$ and 
integrating the path integral over $Q(t)$, 
we have the following 
expression~\cite{Feldbrugge:2017kzv},
\begin{equation}\label{G-amplitude3}
G(q_{f};q_{i}) =  \sqrt{\frac{3\pi i}{2\hbar}}\int_\mathcal{C} \frac{\mathrm{d} N}{N^{1/2}} 
\exp \left(\frac{i S_0[N]}{\hbar}\right),
\end{equation}
where $S_0[N]$ is the semi-classical action,
\begin{align}\label{G-semi-classical}
&S_0[N]  =2\pi^2\int_0^1 \mathrm{d}t 
\left( -\frac{3}{4 N}\dot{q}_s^2 + 3KN - N \Lambda q_s \right) 
\nonumber \\
&= 2\pi^2 \left\{ 
 \frac{N^3\Lambda^2}{36} + N \left( -\frac{\Lambda (q_i+q_f)}{2} +3K \right) 
 -\frac{3(q_f-q_i)^2}{4N} \right\}\,.
\end{align}

We wil integrate the lapse integral 
along the Lefschetz thimbles ${\cal J}_\sigma$ as follows,
\begin{align}\label{G-PL-integral}
G(q_{1};q_{0}) & =  \sum_\sigma n_\sigma \sqrt{\frac{3\pi i}{2\hbar }}
\int_{{\cal J}_\sigma} \frac{\mathrm{d}N}{N^{1/2}} \exp\left(\frac{i S_0[N]}{\hbar}\right).
\end{align}
The lapse integral~\eqref{G-PL-integral} can be estimated based on 
the four saddle points $N_s$,
\begin{equation}\label{G-saddle-lapse-L}
N_s=a_1\frac{3}{\Lambda}
\left[\left(\frac{\Lambda}{3}q_{0}-K\right)^{1/2}
+a_2\left(\frac{\Lambda}{3}q_{1}-K\right)^{1/2}\right],
\end{equation}
where $a_1,a_2\in\{-1,1\}$.
The four saddle points~\eqref{G-saddle-lapse-L} correspond to the intersection
of the Lefschetz thimble $\cal J_\sigma$ and steepest ascent 
paths $\cal K_\sigma$ 
where $\textrm{Re}\left[iS_0\left( N \right)\right]$
decreases and increases monotonically on $\cal J_\sigma$ and $\cal K_\sigma$,
and $n_\sigma$ is the intersection number.
The saddle-point action $S_0[N_s]$ is given by
\begin{align}\label{G-saddle-action-L}
S_0[N_s] = -a_1 \frac{12\pi^2}{\Lambda}  \left[ \left(\frac{\Lambda}{3}q_{0}-K\right)^{3/2} +a_2 \left(\frac{\Lambda}{3}q_{1}-K\right)^{3/2} \right]\,.
\end{align}

By imposing the condition $q_0 = 0$, $K=1$~\cite{Hartle:1983ai}, 
and assuming $q_1{\Lambda} > {3}$,
the gravitational amplitude~\eqref{G-amplitude3} 
corresponds to the cosmological wave function created from nothing $q=0$.
Based on the Lorentzian Picard-Lefschetz method,
Feldbrugge $\textrm{et al.}$~\cite{Feldbrugge:2017kzv} showed that 
the gravitational transition amplitude~\eqref{G-amplitude3} 
by perfuming the integral over a contour in $(0,\infty)$
reduces to the Vilenkin's tunneling wave function~\cite{Vilenkin:1982de}.
On other hand, Diaz Dorronsoro $\textrm{et al.}$~\cite{DiazDorronsoro:2017hti}
reconsidered the gravitational amplitude~\eqref{G-PL-integral}
by integrating the lapse over a different
contour in $(-\infty,\infty)$
and showed that the gravitational amplitude~\eqref{G-PL-integral} reduces to be 
the Hartle-Hawking's no boundary wave function~\cite{Hartle:1983ai}.

%%%%%%%%%%%%%%%%%%%%%%%%%%%%%%%%%%%%%%%%%%%%%%%%
%%%%%%%%%%%%%%%%%%%%%%%%%%%%%%%%%%%%%%%%%%%%%%%%
\begin{figure}[t]
	\centering
	\includegraphics[scale=0.6]{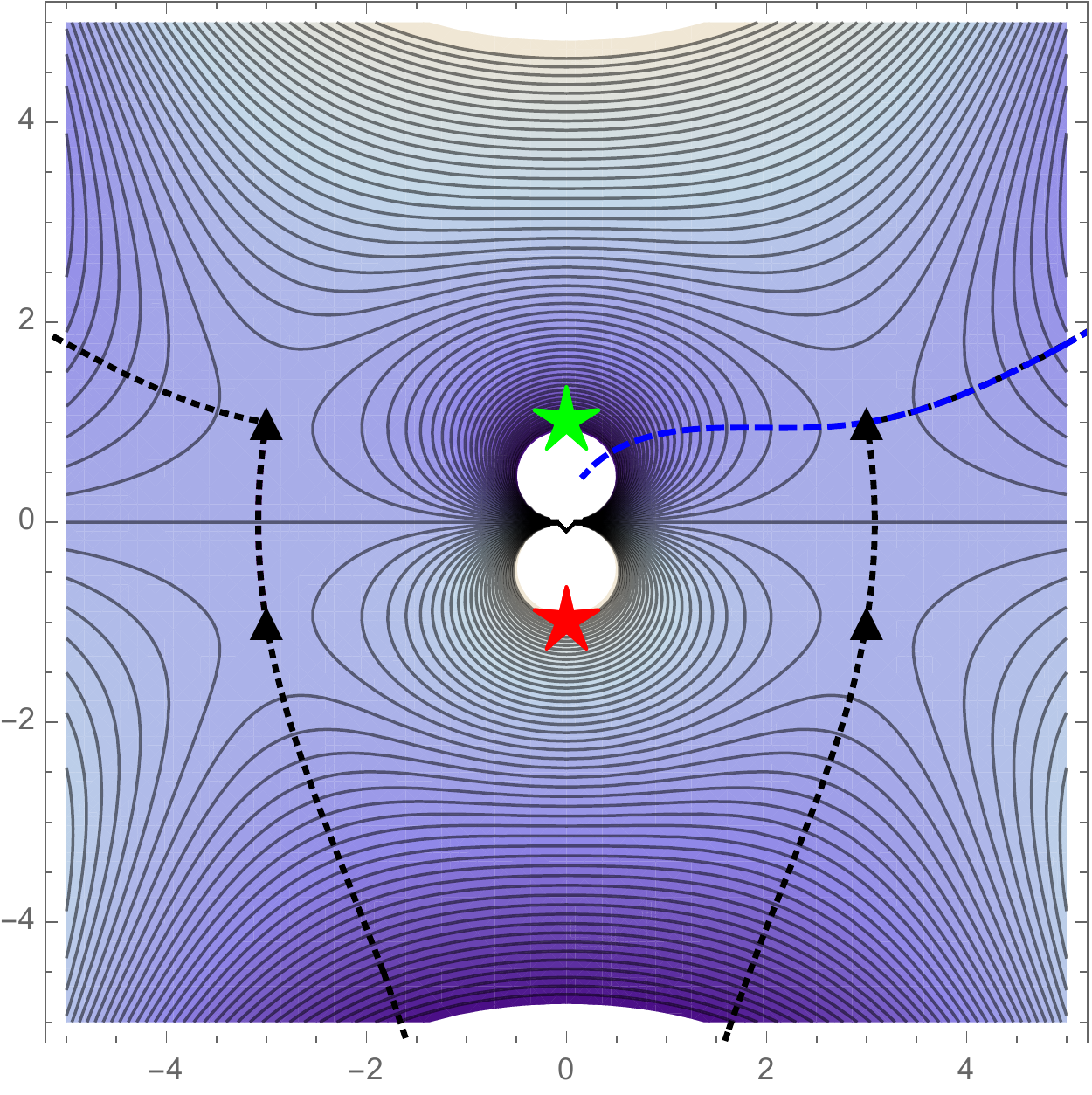}
	\includegraphics[scale=0.6]{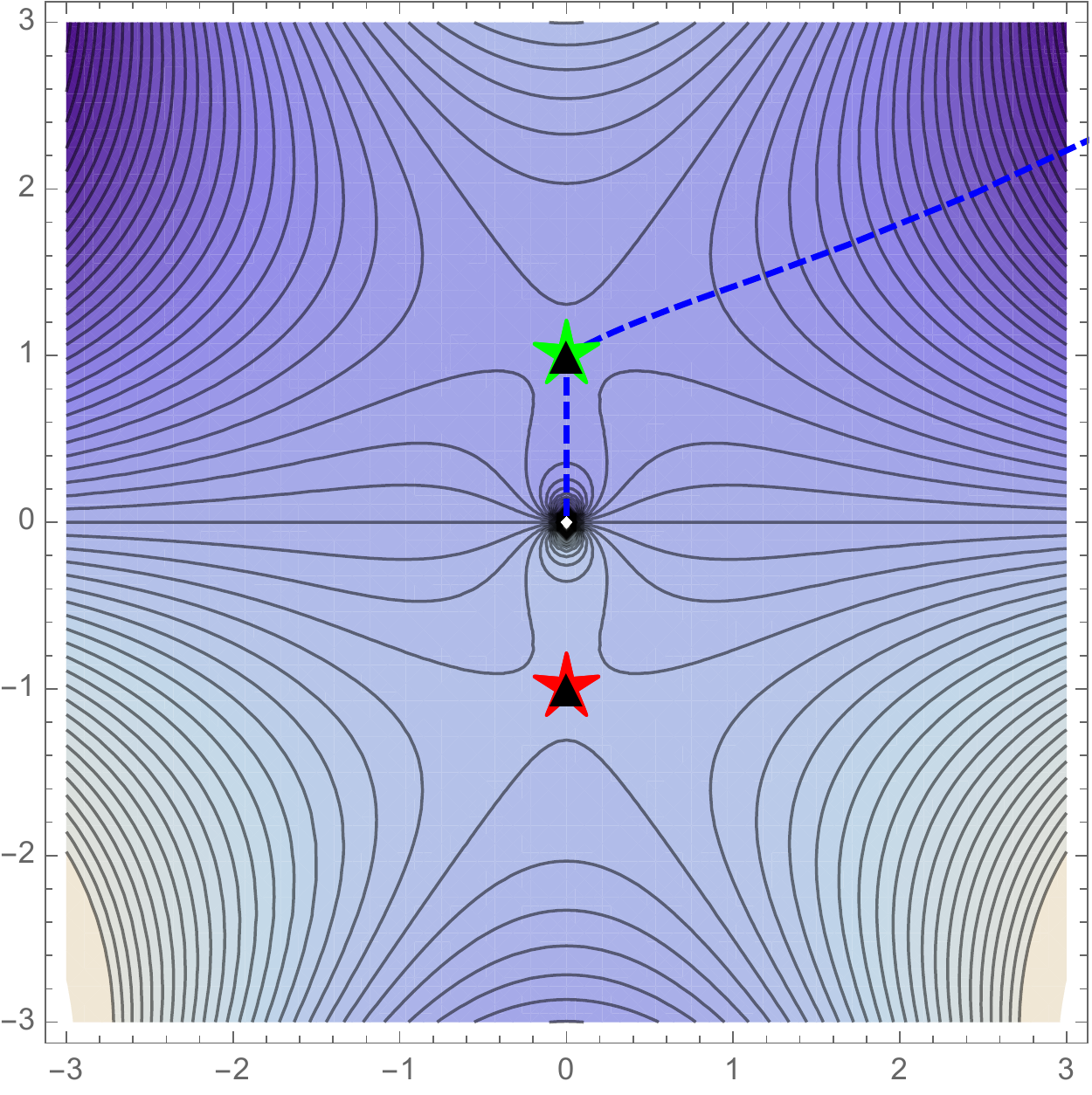}
	\caption{
	In the top figure we set $q_1=10$ and $\Lambda=3$.
	The blue dashed line shows the 
	Lefschetz thimbles on $N=(0,\infty)$ whereas the black dashed line 
	shows the Lefschetz thimbles on $N=(-\infty,\infty)$. The red or green star 
	express the standard or inverted Wick rotation, $N=+i,-i$.
On the other hand, in the bottom figure
we take $q_1=1$ and $\Lambda=3$
and this figure shows $N_s$ coincides with the anti-Euclidean saddle point $N=+i$.
Thus, the tunneling wave function based on the Lorentzian Picard-Lefschetz Formulation
consistent with the Linde wave function~\cite{Linde:1983mx}.}
\label{fig:G-Picard-Lefschetz}
\end{figure}
%%%%%%%%%%%%%%%%%%%%%%%%%%%%%%%%%%%%%%%%%%%%%%%%
%%%%%%%%%%%%%%%%%%%%%%%%%%%%%%%%%%%%%%%%%%%%%%%%

The top figure of 
Fig.~\ref{fig:G-Picard-Lefschetz} discribes $\textrm{Re}\left[iS_0\left( N \right)\right]$
in the complex plane where we set $V_0=3$, $\Lambda=3$ and $q_1=10$.
In the top figure of 
Fig.~\ref{fig:G-Picard-Lefschetz} the Picard-Lefschetz theory says that 
only one Lefschetz thimble can be chosen
in the integration domain $N=(0,\infty)$. On the other hand, 
integrating the complex lapse integral along $N=(-\infty,\infty)$
all four saddle points can contribute the gravitational amplitude~\eqref{G-PL-integral}
although as pointed out in Ref~\cite{Feldbrugge:2017mbc} one can choose 
a different contour in $N=(-\infty,\infty)$.
Hence, the gravitational amplitude~\eqref{G-PL-integral} 
leads to the tunneling and no-boundary wave function,
\begin{align}\label{}
&G_T(q_{1}) \approx c_1 e^{-\frac{12 \pi^2}{\hbar \Lambda} \, -i 4\pi^2 \sqrt{\frac{\Lambda}{3}} (q_1 - 3/\Lambda)^{3/2}/\hbar}\,,\\
&G_{HH}(q_{1}) \approx  c_1 e^{-\frac{12 \pi^2}{\hbar \Lambda} \, -i 4\pi^2 \sqrt{\frac{\Lambda}{3}} (q_1 - 3/\Lambda)^{3/2}/\hbar}\nonumber \\
&+c_2 e^{-\frac{12 \pi^2}{\hbar \Lambda} \, +i 4\pi^2 \sqrt{\frac{\Lambda}{3}} (q_1 - 3/\Lambda)^{3/2}/\hbar} \nonumber \\
&+c_3 e^{+\frac{12 \pi^2}{\hbar \Lambda} \, +i 4\pi^2 \sqrt{\frac{\Lambda}{3}} (q_1 - 3/\Lambda)^{3/2}/\hbar}\nonumber \\
&+c_4 e^{+\frac{12 \pi^2}{\hbar \Lambda} \, -i 4\pi^2 \sqrt{\frac{\Lambda}{3}} (q_1 - 3/\Lambda)^{3/2}/\hbar}\,,
\end{align}
where $c_{1,2,3}$ include the functional determinants and prefactors.

Let us consider the gravitational amplitude~\eqref{G-amplitude2} again but 
take a different semi-classical approximation to the gravitational action.
In the previous discussion, the gravitational action was semi-classically approximated by
the classical solutions of the equation of motion, but we can 
approximate the gravitational action
by utilizing the constraint equation as discussed in Section~\ref{sec:Euclidean-path-integral}.
Thus, by solving the constraint equation
for $\dot{q}$ and substituting in the action,
we can obtain the semi-classical gravitational action as follows~\cite{Brown:1990iv,Vilenkin:2018dch,deAlwis:2018sec},
\begin{align}
\begin{split}
& S_0 [q_{1};q_{0}]=  4\pi^2\int_{t_0}^{t_1}Ndt\left(3K-\Lambda q\right)\\
 &  = \pm\,2\sqrt{3}\pi^2\int_{q_{0}}^{q_{1}}dq\sqrt{\Lambda q- 3K}\\
 &  = \pm \frac{12\pi^2}{\Lambda}  \left[ \left(\frac{\Lambda}{3}q_{0}-K\right)^{3/2} - \left(\frac{\Lambda}{3}q_{1}-K\right)^{3/2} \right], 
\end{split}
\end{align}
which cancels the lapse $N$ contribution~\cite{deAlwis:2018sec} 
and corresponds to
the semi-classical action under the WKB
approximation. In fact, in the leading order
the WKB wave function is given by
\begin{equation}
\Psi[q_{1};q_{0}]\approx C \cdot
\exp \left[{\pm\frac{i\,2\sqrt{3}\pi^2}{\hbar}\int_{q_{0}}^{q_{1}}dq\sqrt{\Lambda q- 3K}}\right]\,.
\end{equation}
where the exponent of the above WKB wave function
agrees with the two saddle-point actions~\eqref{G-saddle-action-L}.

From here, let us discuss the relation between the wave function in 
the Lorentzian Picard-Lefschetz Formulation and the
Hartle-Hawking or Linde wave function given by the Euclidean path integral.
We can easily see that setting $N=-i$
the Lorentzian path integral  
corresponds to the Euclidean path integral
since $S[q,N]$ is given by the Wick-rotation $t\rightarrow Nt$ and $S_E[q]$ 
is given by $t\rightarrow -i\tau$.
For simplicity, considering the original gravitational action $S[a]$
and redefining the time $t=\tau$,
the Euclidean action $S_E[a]$ for the closed universe with $K=1$ is given by
\begin{align}
\begin{split}
&iS[a,-i] \equiv -S_E[a]\\
&=2\pi^2 \int_{0}^{\tau} \mathrm{d}\tau \left( 3 a
\left(\frac{d a}{d\tau}\right)^2 + 3a - 3a ^3H^2 \right) ,
\label{Euclidean-action}
\end{split}
\end{align}
where $H^2=\Lambda/3$ and $S_E[a]$  is negative for small $a$.
From \eqref{Euclidean-action} we obtain the Euclidean constraint equation 
and the equations of motion,
\begin{align}
\left(\frac{d a}{d\tau}\right)^2 - 1 + a^2H^2=0, \quad
\left(\frac{d^2 a}{d\tau^2}\right) = -aH^2.
\end{align}
Hence, we obtain the Euclidean de Sitter solution 
$a(\tau)= H^{-1}\sin H\tau$ with the initial condition $a(0)=0$.
Note that in the saddle point method of the Euclidean path integral, 
the classical solution of the saddle point is not correct 
unless not only the equations of motion but also 
the constraint equation are satisfied.

By using the classical saddle-point solution we can 
evaluate the Euclidean action
under the saddle-point approximation,
\begin{align}
S_E[a] &=- 2\pi^2 \int_{0}^{\pi /2H} \mathrm{d}\tau \left( 3 a
\left(\frac{d a}{d\tau}\right)^2 + 3a - 3a ^3H^2 \right)\notag  \\
&= - \frac{12\pi^2}{\Lambda}\ .
\end{align}
Thus, we have the Hartle-Hawking wave function,
\begin{equation}
\Psi_{HH}(a)\sim \exp\left(-S_E[a]\right) \sim
\exp\left(+\frac{12\pi^2}{\Lambda}\right) .
\end{equation}
which diverges the probability and disfavors the inflationary cosmology
if we replace the cosmological constant $\Lambda$ with the 
scalar potential $V(\phi)$.
On the other hand, to suppress the exponential probability and 
get the cosmological wave function of the ground state, 
Linde~\cite{Linde:1983mx} proposed 
the wave function utilizing the anti-Wick rotation $\tau = -i t$,
\begin{equation}
\Psi_{L}(a)\sim \exp\left(+S_E[a]\right) \sim\exp
\left(-\frac{12\pi^2}{\Lambda}\right) ,
\end{equation}

Now, we can easily show that the wave function from 
the Lorentzian Picard-Lefschetz Formulation
and WKB method includes the Hartle-Hawking or 
Linde wave function given by the saddle point method of the Euclidean path integral. 
As already mentioned, the de Sitter solution 
$a(\tau)= H^{-1}\sin H\tau$  is a saddle-point solution 
with zero energy, and the final scale factor is $a(\tau)= H^{-1}$. 
In other words, the saddle-point method of Euclidean path integral 
only gives the transition amplitude from the entry to the exit of the potential 
in the quantum tunneling. 
In the Lorentzian Picard-Lefschetz Formulation
and WKB approximation method, such a transition amplitude 
can be given by $\Lambda q_1- 3K=0$. Thus, 
we get the corresponding transition amplitude,
\begin{align}
\Psi[q_{1}]&\approx  \sqrt{\frac{3\pi i}{2\hbar}}\int_{0,-\infty}^{+\infty} \frac{\mathrm{d} N}{N^{1/2}} 
\exp \left(\frac{i S_0[N]}{\hbar}\right)\notag \\
&\approx  \exp \left(\pm \frac{12\pi^2}{\Lambda}\right) \,.
\end{align}
where we note that the four saddle points~\eqref{G-saddle-lapse-L} of the 
semi-classical action $S_0[ N]$ in 
the Lorentzian Picard-Lefschetz method converges on
the two saddle points $N_s=\pm i$.
In the bottom figure of 
Fig.~\ref{fig:G-Picard-Lefschetz} we show the correspondence.

\medskip
%%%%%%%%%%%%%%%%%%%%%%%%%%%%%%%%%%%%
%%%%%%%%%%%%%%%%%%%%%%%%%%%%%%%%%%%%
\section{Quantum tunneling with Lorentzian instanton} 
\label{sec:Lorentzian-instanton} 
%%%%%%%%%%%%%%%%%%%%%%%%%%%%%%%%%%%%
%%%%%%%%%%%%%%%%%%%%%%%%%%%%%%%%%%%%
In this section, we will introduce a new 
instanton method based on the previous discussions.
As discussed in Section~\ref{sec:Euclidean-path-integral}, 
the Euclidean saddle-point action 
corresponding to the exponent of the WKB wave function
does not consist only of the simple 
solutions of the equation of motion. The solutions of the Euclidean equation of motion 
must satisfy the constraint equation 
\eqref{eq:Neqn} in the Euclidean form. 
The Lorentzian Picard-Lefschetz formulation~\cite{Feldbrugge:2017kzv}
constructs the semi-classical transition amplitude by finding saddle points on the lapse 
integration which implies $\delta S[x,N]/\delta N=0$. 
Since $\delta S[x,N]/\delta N=0$ corresponds to the constraint equation 
\eqref{eq:Neqn}, 
 the transition amplitude becomes consistent with the WKB approximation 
 for wave-function.
Hence, we can expect to obtain the saddle-point action corresponding to the WKB approximation by finding the Lorentzian solution from the constraint equation 
\eqref{eq:Neqn} and substituting it into the Lorentzian action. 
We will now briefly introduce the method and call it 
$\it Lorentzian\ instanton\ formulation$.

Let us write the Lorentzian
path integral including lapse function $N_L$ again,
\begin{align}\label{QM-action2}
K(x_{f};x_{i})&=\int_{x_{i}}^{x_{f}} 
\mathcal{D}x ~ \exp \left(\frac{i S[x]}{\hbar}\right), \\
S[x]&=\int_{t_i}^{t_f}dt N_L
\left( \frac{\dot{x}^{2}}{2N_L^2}-V(x)+E\right),
\end{align}
which fixes the lapse $N_L$ and does not integrate.
From \eqref{QM-action2} we derive the constraint equation 
and the equations of motion,
\begin{align}
\delta S[x]/\delta{N_L}& =  0\ \Longrightarrow \ 
\frac{\dot{x}^{2}}{2}+N_L^{2}V(x)=N_L^{2}E\label{eq:Neqn2},\\
\delta S[x]/\delta{x} & =  0 \ \Longrightarrow \ \ddot{x}=-N_L^{2}V'(x)\label{eq:xeqn2}.
\end{align}

As is well known in analytical mechanics, 
the equation of motion~\eqref{eq:Neqn2}  is obtained by differentiating the constraint equation~\eqref{eq:xeqn2}.
Thus, only the constraint equation is considered.
Solving the constraint equation~\eqref{eq:xeqn2} for $x$
with the initial condition $x(t_i)=x_0$,
we get one semi-classical solution. Then, we impose the final condition $x(t_f)=x_1$
on the solution and determine the lapse function $N_L$ on the complex path.
Thus, we can get the Lorentzian real-time solution even for quantum tunneling
and construct the path integral~\eqref{QM-action2} from the Lorentzian 
classical solution as well as the instanton method based on the Euclidean path integral.

From here, we will show that this formulation is consistent with 
Lorentzian Picard-Lefschetz formulation~\eqref{PL-integral} and WKB approximation
for the wave function.
For simplicity, let us consider the linear potential $V=V_{0}-\Lambda x$
and assume $t_i=0$ and $t_f=1$.
The constraint equation~\eqref{eq:xeqn2} with the initial condition $x(t_i=0)=x_0$
gives the following classical solution,
\begin{equation}\label{x-semisol2}
x_{L}(t)=x_0+\frac{\Lambda }{2}N_L^2 t^2\pm N_Lt \sqrt{2E-2V_0+2 \Lambda x_0}.
\end{equation}
By imposing the final condition $x(t_f=1)=x_1$
on the solution we can determine the lapse function $N_L$ and obtain, 
\begin{align}
&N_{L}=\mp\frac{\sqrt{2}}{\Lambda}
\left[(\Lambda x_{0}+E-V_{0})^{1/2}
\pm(\Lambda x_{1}+E-V_{0})^{1/2}\right] ,\notag \\
&x_{L}(t)=x_0+
\left[(\Lambda x_{0}+E-V_{0})^{1/2}
\pm(\Lambda x_{1}+E-V_{0})^{1/2}\right]^2 t^2\nonumber \\
&- \frac{\sqrt{2}\, t}{\Lambda}
\left[(\Lambda x_{0}+E-V_{0})^{1/2}
\pm(\Lambda x_{1}+E-V_{0})^{1/2}\right]\notag \\
&\times \sqrt{2E-2V_0+2 \Lambda x_0}.\label{x-semisol3}
\end{align}
where $N_{L}$ corresponds to the four saddle points~\eqref{saddle-lapse-L} of the 
Lorentzian Picard-Lefschetz formulation and $x_{L}(t)$ is given by
imposing $N_{L}$ to the Lorentzian solution~\eqref{x-semisol2}.
Note that plus sign solution in Eq.~\eqref{x-semisol3}
is non-trivial since it is non-zero in the limit $x_{1} \rightarrow x_{0}$.
Here, the degree of freedom of lapse is uniquely determined. 
Thus, the semi-classical action $S[x_L]$ is given by
\begin{align}
S[x_L]&=\pm\frac{2\sqrt{2}}{3\Lambda}  \left[\left(\Lambda x_{0}+E-V_{0}\right)^{3/2} \pm
 \left(\Lambda x_{1}+E-V_{0}\right)^{3/2} \right].
 \end{align}
which is consistent with 
the semi-classical action for the WKB wave function and 
saddle-point action~\eqref{saddle-action-L} of the Lorentzian Picard-Lefschetz formulation.
By developing the saddle point method 
where $x=x_L+\delta x$ is decomposed as the 
Lorentzian classical solutions and the fluctuation
around them, 
the Lorentzian transition amplitude can be approximately given by
%%%%%%%%%%%%%%%%%%%%%%%%%%%%%%%%%
\footnote{
\begin{widetext}
~When $V''(x_L)$ is non-zero,
as the usual instanton in QFT~\cite{Callan:1977pt},
we can expand the fluctuation and get the following expression, 
\begin{align*}
\begin{split}
K(x_{f};x_{i})
&\simeq \exp \left(\frac{i S[x_L]}{\hbar}\right)
\int_{\delta x(0)=0}^{\delta x(1)=0}\mathcal{D}\delta x\, \\
&\times e^{\left(\frac{i}{2\hbar}
\int_{0}^{1}dt N_L
\left( \frac{d^{2}}{N_L^2dt^2}-V''(x_L)\right)\delta x^2\right)}\\
&\simeq \exp \left(\frac{i S[x_L]}{\hbar}\right)
\int_{-\infty}^{\infty}dc_n\, \exp \left(\frac{-1}{2\hbar iN_L}\sum_{n=0}^{\infty}c_n^2\lambda_n\right)\\
&\simeq \exp \left(\frac{i S[x_L]}{\hbar}\right)
\prod_{n}\sqrt{\frac{2\pi \hbar i N_L}{\lambda_n}},
\end{split}
\end{align*}
where $\lambda_n$ are the eigenvalues and we omitted the normalization.
For the Euclidean path integral where $N_L=-i$, 
when all eigenvalues in a saddle point solution are positive, 
the fluctuation around the saddle point increases the action and 
the saddle point approximation is correct. On the other hand, the 
negative eigenvalues reduce the action, 
and such a solution would be discarded~\cite{Coleman:1987rm}. 
A similar argument can be applied here.
\end{widetext}}
%%%%%%%%%%%%%%%%%%%%%%%%%%%%%%%%%
\begin{align}\label{Lorentzian-method2}
\begin{split}
&K(x_{1};x_{0})\simeq \exp \left(\frac{i S[x_L]}{\hbar}\right)\int_{\delta x(0)=0}^{\delta x(1)=0}\mathcal{D}\delta x\, 
e^{\left(\frac{i}{2\hbar}
\frac{\delta^2S[x]}
{\delta x^2}\Bigr|_{x=x_L}\delta x^2\right)}\\
&\simeq \exp \left(\frac{i S[x_L]}{\hbar}\right)
\int_{\delta x(0)=0}^{\delta x(1)=0}\mathcal{D}\delta x\, 
e^{\left(\frac{i}{2\hbar N_L}
\int_{0}^{1}dt 
\left( \delta \dot{x}^2
-V''(x_L)\delta x^2\right)\right)}\\
&\simeq \sqrt{\frac{1}{2\pi i\hbar N_L}}\exp \left(\frac{i S[x_L]}{\hbar}\right),
\end{split}
\end{align}
where $V''(x_L)$ is zero for the linear potential.
Note that this method still has a problem 
how to select the correct Lorentzian solutions.
Let us compare this result with the real-time path integral
for the linear potential. Following Feynman's 
famous textbook~\cite{feynman2010quantum}, 
the real-time transition amplitude is approximately given by
\begin{align}\label{real-time-path}
\begin{split}
K(x_{1};x_{0})&\simeq \sqrt{\frac{1}{2\pi i\hbar }}
e^{\left[\frac{i}{\hbar}\left\{\frac{(x_{1}-x_{0})^{2}}{2}
+\frac{\Lambda(x_{1}+x_{0})}{2}-\frac{\Lambda^{2}}{24}\right\}\right]},
\end{split}
\end{align}
which does not agree with the Lorentzian formulation~\eqref{Lorentzian-method2}.  
However, in this saddle point approximation the classical path solution 
does not respect the constraint equation~\eqref{eq:xeqn} 
and if we explicitly write the energy of the system
and usually think about that, $x_{1,0}$ is restricted by $E$ as classical mechanics.
By imposing the constraint~\eqref{eq:xeqn} on the real-time classical solution
and setting $x_0=0$ and $V_0=E=0$ for simplicity, 
the real-time transition amplitude~\eqref{real-time-path} agrees 
with the Lorentzian amplitude~\eqref{Lorentzian-method2},
\begin{align}
K(x_{1};0)
\simeq \sqrt{\frac{1}{2\pi i\hbar }}\exp \left(\frac{4ix_1^2}{3\hbar}\right).
\end{align}
As discussed in Section~\ref{sec:Euclidean-path-integral}, 
the Lorentzian transition amplitudes, 
which approximate the saddle point from the constraint equation~\eqref{eq:xeqn}
in correspondence with the WKB analysis, 
would be more accurate than the real-time amplitude~\eqref{real-time-path}
if the energy of the system is fixed.

\bigskip
%%%%%%%%%%%%%%%%%%%%%%%%%%%%%%%%%%%%
%%%%%%%%%%%%%%%%%%%%%%%%%%%%%%%%%%%%
\section{Conclusion}
\label{sec:Conclusion} 
%%%%%%%%%%%%%%%%%%%%%%%%%%%%%%%%%%%%
%%%%%%%%%%%%%%%%%%%%%%%%%%%%%%%%%%%%

In this paper, we have analyzed the tunneling transition 
amplitude for QM using the Lorentzian 
Picard-Lefschetz formulation~\eqref{PL-integral} for QM 
and compare it with the WKB analysis of the conventional Schr\"{o}dinger
equation. In the literature~\cite{Feldbrugge:2017kzv,DiazDorronsoro:2017hti,
Feldbrugge:2017fcc,Feldbrugge:2017mbc,
Halliwell:1988ik,Halliwell:1989vu,Halliwell:1990tu,
Brown:1990iv,Vilenkin:2018dch,deAlwis:2018sec} the gravitational transition 
amplitude using the method of steepest descents  
and the Picard-Lefschetz theory corresponds to the WKB solution 
of the Wheeler-DeWitt equation.
We have shown that they are agreement for 
the linear, harmonic oscillator, inverted harmonic oscillator, and double well models
in QM.
The two saddle points of the Lorentzian 
Picard-Lefschetz formulation~\eqref{PL-integral}
corresponds to the exponents of the WKB wave function whereas  
the others are conjugate.
These results suggest that the Lorentzian 
Picard-Lefschetz formulation 
is consistent with the WKB analysis of the Schr\"{o}dinger equation for QM.
In Section~\ref{sec:Euclidean-path-integral}
we have argued why the Lorentzian Picard-Lefschetz formulation is consistent with 
the WKB analysis of the conventional Schr\"{o}dinger
equation. Applying the saddle-point method of action 
to satisfy the constraint equation~\eqref{eq:xeqn} leads to 
the correct semiclassical approximation of the path integral. 
In Section~\ref{sec:Lorentzian-gravity}
we have demonstrated that 
the tunneling and no-boundary wave functions derived by 
the Lorentzian Picard-Lefschetz Formulation corresponds to the WKB solution 
of the Wheeler-DeWitt equation.
Finally, in Section~\ref{sec:Lorentzian-instanton} we have provided 
a simpler semi-classical approximation way of the Lorentzian path integral
without integrating the lapse function.

\medskip
%%%%%%%%%%%%%%%%%%%%
\para{Acknowledgements} 
We thank K. Yamamoto for stimulating discussions and valuable comments. 
We also thank S. Kanno, A. Matsumura, and H. Suzuki for helpful discussions.
%%%%%%%%%%%%%%%%%%%%

%%%%%%%%%%%%%%%%%%%%%%%%%%%%%%%%%%%%%%%%%%%%%%%%%%%%%%%%%%%%%%%%%%%%
\nocite{}
\bibliography{arXiv-version}
%%%%%%%%%%%%%%%%%%%%%%%%%%%%%%%%%%%%%%%%%%%%%%%%%%%%%%%%%%%%%%%%%%%%
\end{document}